\documentclass[12pt,manuscript]{aastex}

\newcommand{\teff}{\ensuremath{T_{\rm eff}}}
\newcommand{\logg}{\ensuremath{\log{g}}}

\newcommand{\feh}{[Fe/H]}

\newcommand{\rsun}{\ensuremath{R_\sun}}
\newcommand{\msun}{\ensuremath{M_\sun}}

\newcommand{\rstar}{\ensuremath{R_\star}}
\newcommand{\mstar}{\ensuremath{M_\star}}

\newcommand{\rhostar}{\ensuremath{\rho_\star}}

\newcommand{\rpl}{\ensuremath{R_{\rm P}}}

\newcommand{\rearth}{\ensuremath{R_{\earth}}}

\newcommand{\nCandidatesNewTotal}{1091}

\newcommand{\nearth}{196}
\newcommand{\nsuperearth}{416}
\newcommand{\nneptune}{421}
\newcommand{\njupiter}{41}
\newcommand{\nlarger}{17}
\newcommand{\nexcluded}{10}
\newcommand{\nexcludedsingles}{2}

\newcommand{\febSmallerThanNeptune}{\ensuremath{73\%}}
\newcommand{\newSmallerThanNeptune}{\ensuremath{91\%}}

\newcommand{\nCandidatesTotal}{2321}
\newcommand{\nStarsTotal}{1790}

\newcommand{\gainSmall}{\ensuremath{197\%}}
\newcommand{\gainLarge}{\ensuremath{52\%}}
\newcommand{\gainShort}{\ensuremath{85\%}}
\newcommand{\gainLong}{\ensuremath{123\%}}

\newcommand{\fracSmallShort}{2.8}
\newcommand{\fracSmallLong}{7.2}
\newcommand{\fracLarge}{1.6}

\newcommand{\hzTotal}{46}
\newcommand{\hzSuperEarthSize}{9}
\newcommand{\hzEarthSize}{1}

\newcommand{\hzTotalFeb}{22}

\newcommand{\nTwo}{245}
\newcommand{\nThree}{84}
\newcommand{\nFour}{27}
\newcommand{\nFive}{8}

\newcommand{\nStarsWithMulti}{365}
\newcommand{\nMultiCandidates}{896}

\def\note #1]{{\bf #1]}}

\newcommand{\ek}{\emph{Kepler}}

\shortauthors{Batalha et al.}
\shorttitle{Kepler Planet Candidates. III.}

\begin{document}

\bibliographystyle{apj}

\title{Planetary Candidates Observed by \emph{Kepler} \\ III: Analysis of the First 16 Months of Data}

\author{
Natalie~M.~Batalha$^{1}$,
Jason~F.~Rowe$^{2}$,
Stephen~T.~Bryson$^{3}$,
Thomas~Barclay$^{4}$,
Christopher~J.~Burke$^{2}$,
Douglas~A.~Caldwell$^{2}$,
Jessie~L.~Christiansen$^{3}$,
Fergal~Mullally$^{2}$,
Susan~E.~Thompson$^{2}$,
Timothy~M.~Brown$^{5}$,
Andrea~K.~Dupree$^{6}$,
Daniel~C.~Fabrycky$^{7}$,
Eric~B.~Ford$^{8}$,
Jonathan~J.~Fortney$^{7}$,
Ronald~L.~Gilliland$^{9}$,
Howard~Isaacson$^{10}$,
David~W.~Latham$^{6}$,
Geoffrey~W.~Marcy$^{10}$,
Samuel~N.~Quinn$^{6,29}$,
Darin~Ragozzine$^{6}$,
Avi~Shporer$^{5}$,
William~J.~Borucki$^{3}$,
David~R.~Ciardi$^{11}$,
Thomas~N.~Gautier III$^{12}$,
Michael~R.~Haas$^{3}$,
Jon~M.~Jenkins$^{2}$,
David~G.~Koch$^{3}$,
Jack~J.~Lissauer$^{3}$,
William~Rapin$^{3}$,
Gibor~S.~Basri$^{10}$,
Alan~P.~Boss$^{13}$,
Lars~A.~Buchhave$^{14,15}$,
David~Charbonneau$^{6}$,
Joergen~Christensen-Dalsgaard$^{16}$,
Bruce~D.~Clarke$^{12}$,
William~D.~Cochran$^{17}$,
Brice-Olivier~Demory$^{18}$,
Edna~Devore$^{19}$,
Gilbert~A.~Esquerdo$^{6}$,
Mark~Everett$^{20}$,
Francois~Fressin$^{6}$,
John~C.~Geary$^{6}$,
Forrest~R.~Girouard$^{21}$,
Alan~Gould$^{22}$,
Jennifer~R.~Hall$^{21}$,
Matthew~J.~Holman$^{6}$,
Andrew~W.~Howard$^{10}$,
Steve~B.~Howell$^{3}$,
Khadeejah~A.~Ibrahim$^{21}$,
K.~Kinemuchi$^{2}$,
Hans~Kjeldsen$^{16}$,
Todd~C.~Klaus$^{21}$,
Jie~Li$^{2}$,
Philip~W.~Lucas$^{23}$,
Robert~L.~Morris$^{21}$,
Andrej~Pr\u{s}a$^{24}$,
Elisa~Quintana$^{2}$,
Dwight~T.~Sanderfer$^{21}$,
Dimitar~Sasselov$^{6}$,
Shawn~E.~Seader$^{2}$,
Jeffrey~C.~Smith$^{2}$,
Jason~H.~Steffen$^{25}$
Martin~Still$^{26}$,
Martin~C.~Stumpe$^{2}$,
Jill~C.~Tarter$^{19}$,
Peter~Tenenbaum$^{2}$,
Guillermo~Torres$^{6}$,
Joseph~D.~Twicken$^{2}$,
Kamal~Uddin$^{21}$,
Jeffrey~Van~Cleve$^{2}$,
Lucianne~Walkowicz$^{27}$,
William~F.~Welsh$^{28}$
}
\affil{$^{1}$Department of Physics and Astronomy, San Jose State University, San Jose, CA 95192}
\affil{$^{2}$SETI Institute/NASA Ames Research Center, Moffett Field, CA 94035}
\affil{$^{3}$NASA Ames Research Center, Moffett Field, CA 94035}
\affil{$^{4}$Bay Area Environmental Research Institute/NASA Ames Research Center, Moffett Field, CA 94035}
\affil{$^{5}$Las Cumbres Observatory Global Telescope Network, Goleta, CA 93117}
\affil{$^{6}$Harvard-Smithsonian Center for Astrophysics, 60 Garden Street, Cambridge, MA 02138}
\affil{$^{7}$Department of Astronomy and Astrophysics, University of California, Santa Cruz, CA 95060 USA}
\affil{$^{8}$University of Florida, Gainesville, FL 32611}
\affil{$^{9}$Center for Exoplanets and Habitable Worlds, The Pennsylvania State University, University Park, PA 16802}
\affil{$^{10}$University of California, Berkeley, Berkeley, CA 94720}
\affil{$^{11}$NASA Exoplanet Science Institute/Caltech, Pasadena, CA 91125}
\affil{$^{12}$Jet Propulsion Laboratory/California Institute of Technology, Pasadena, CA 91109}
\affil{$^{13}$Carnegie Institution of Washington, Washington, DC 20015-1305}
\affil{$^{14}$Niels Bohr Institute, University of Copenhagen, DK-2100, Copenhagen, Denmark}
\affil{$^{15}$Centre for Star and Planet Formation, Natural History Museum of Denmark, University of Copenhagen, DK-1350 Copenhagen, Denmark}
\affil{$^{16}$Aarhus University, DK-8000 Aarhus C, Denmark}
\affil{$^{17}$McDonald Observatory, The University of Texas, Austin TX 78712}
\affil{$^{18}$Department of Earth, Atmospheric and Planetary Sciences, Massachusetts Institute of Technology, 77 Massachusetts Ave., Cambridge, MA 02139, USA}
\affil{$^{19}$SETI Institute, Mountain View, CA 94043}
\affil{$^{20}$National Optical Astronomy Observatory, Tucson, AZ 85719}
\affil{$^{21}$Orbital Sciences Corporation/NASA Ames Research Center, Moffett Field, CA 94035}
\affil{$^{22}$Lawrence Hall of Science, Berkeley, CA 94720}
\affil{$^{23}$Centre for Astrophysics, University of Hertfordshire, College Lane, Hatfield AL10 9AB}
\affil{$^{24}$Villanova University, Villanova, PA 19085}
\affil{$^{25}$Fermilab Center for Particle Astrophysics, Batavia, IL 60510}
\affil{$^{26}$Bay Area Environmental Research Institute/NASA Ames Research Center, Moffett Field, CA 94035}
\affil{$^{27}$Department of Astrophysical Sciences, Princeton University, Princeton, NJ 08544}
\affil{$^{28}$San Diego State University, San Diego, CA 92182}
\affil{$^{29}$Department of Physics and Astronomy, Georgia State University, PO Box 4106, Atlanta, Georgia 30302, USA}
\altaffiltext{*}{Correspondences can be addressed to: Natalie.Batalha@sjsu.edu}

\begin{abstract}

New transiting planet candidates are identified in sixteen months (May 2009 - September 2010) of data from the \ek\ spacecraft.  Nearly five thousand periodic transit-like signals are vetted against astrophysical and instrumental false positives yielding \nCandidatesNewTotal\ viable new planet candidates, bringing the total count up to over 2,300.  Improved vetting metrics are employed, contributing to higher catalog reliability. Most notable is the noise-weighted robust averaging of multi-quarter photo-center offsets derived from difference image analysis which identifies likely background eclipsing binaries. Twenty-two months of photometry are used for the purpose of characterizing each of the new candidates.  Ephemerides (transit epoch, $T_{\rm 0}$, and orbital period, $P$) are tabulated as well as the products of light curve modeling: reduced radius ($\rpl/\rstar$), reduced semi-major axis ($d/\rstar$), and impact parameter ($b$). The largest fractional increases are seen for the smallest planet candidates (\gainSmall\ for candidates smaller than $2\rearth$ compared to \gainLarge\ for candidates larger than $2\rearth$) and those at longer orbital periods (\gainLong\ for candidates outside of 50~day orbits versus \gainShort\ for candidates inside of 50~day orbits).  The gains are larger than expected from increasing the observing window from thirteen months (Quarter 1-- Quarter 5) to sixteen months (Quarter 1 -- Quarter 6).  This demonstrates the benefit of continued development of pipeline analysis software.   The fraction of all host stars with multiple candidates has grown from 17\%\ to 20\%, and the paucity of short-period giant planets in multiple systems is still evident.  The progression toward smaller planets at longer orbital periods with each new catalog release suggests that Earth-size planets in the Habitable Zone are forthcoming if, indeed, such planets are abundant. 


\end{abstract}



\section{Introduction}
\label{sec:intro}

Since initiating science operations in May, 2009, \ek\ has produced two catalogs of transiting planet candidates.  The first, released in June 2010, contains 312 candidates identified in the first 43 days of \ek\ data \citep{juneCatalog} and is hereafter referred to as B10.  The second, released in February 2011, is a cumulative catalog containing 1,235 candidates identified in the first 13 months (Quarters 1 through 5)\footnote{Quarters are defined by a requirement to roll the spacecraft 90$^\circ$ about its axis to keep the solar arrays illuminated and the focal-plane radiator pointed away from the Sun. All but the first quarter are approximately 93 days in duration.  In Quarter 1, the spacecraft operated in science mode for 33 days.} of data \citep{febCatalog}.  This cumulative catalog is hereafter referred to as B11.  Over 60 candidates from the B11 catalog have been confirmed, including many of {\it Kepler's} milestone discoveries:  the mission's first rocky planet, Kepler-10b \citep{kepler10}, the six transiting-planet system, Kepler-11 \citep{kepler11}, the first circumbinary planet, Kepler-16ABb \citep{kepler16}, the 2.38 \rearth\ planet in the Habitable Zone, Kepler-22b \citep{kepler22}, and the mission's first Earth-size planets, Kepler-20 e \& f \citep{kepler20}.

In this contribution, we present new planet candidates identified from the analysis of 16 months of data (Quarters 1 through 6).  The analysis was motivated by the availability of the SOC 7.0 pipeline in the summer of 2011.  Though only one more quarter of data was added, it was the first time that multiple quarters were stitched together to produce one continuous time-series as part of {\it Kepler's} pipeline transit search module (TPS).  As such, we expected large gains in the detection of new candidates.

We describe the results of this effort -- the data (Section~\ref{sec:observations}), the procedures that sort transit-like signals coming out of the pipeline into viable planet candidates (Section~\ref{sec:tce}), and the subsequent vetting criteria that lead to increased catalog reliability (Section~\ref{sec:vetting}).  We describe the characterization of the planet candidates (Section~\ref{sec:planetProps}) that begins with transit light curve modeling (Section~\ref{sec:modeling}) and ultimately requires detailed knowledge of the stellar properties.  An effort was made to improve upon the stellar properties from the \ek\ Input Catalog \citep{kic} by utilizing theoretical evolutionary tracks as described in Section~\ref{sec:starProps}.  We examine the distributions of the resulting planet properties (Section~\ref{sec:distributions}) and take a collective look at the progress to date as we work towards the identification of Earth-size planets in the habitable zone. We compare the observed gains to those predicted by way of adding three months of data (Section~\ref{sec:gains}). The new multiple transiting planet systems are briefly described, as are the candidates in the Habitable Zone.  Finally, in Appendix~\ref{sec:bigCatalog}, we provide a cumulative table of planet candidates containing the characteristics of the new candidates as well as updated characteristics of the candidates in the B11 catalog computed using the same data used herein.

\section{Observations}
\label{sec:observations}

The data employed for transit identification were acquired between 2009 May 13 00:15 UTC and 2010 Sep 22 19:03 UTC (Q1 through Q6).  Over 190,000 stars were observed at some time during this period.  Of these, only 127,816 were observed every single quarter.  Therefore, it should not be assumed that every star tabulated herein was observed continuously for the 6 quarter period.  While this is a reasonable assumption for the Feb 2011 catalog (906, or 91\% of the 997 stars identified as planet hosts were observed all five quarters), it is not for the population of new candidates presented here, where only 704 (76\%) of the 926 unique stars identified as planet hosts were observed all six quarters.  The last column of Table~\ref{tab:starProps} presents a string of six integers, each indicating if the target was (one) or was not (zero) observed during the quarter in question (ordered one through six, from left to right).  The 4$^{\rm th}$ integer(corresponding to Quarter 4) can also assume a value of 2, indicating targets located on CCD Module 3 during Quarter 4.  Module 3 failed at 17:52 UTC on 9 January 2010 and never recovered.  Targets located on that module were observed for a shorter time period (see below).  The start and stop times for each quarter are listed in Table~\ref{tab:observations}.  The loss of Module 3 implies that approximately $19$\%\ of {\it Kepler's} targets will be observed three out of four quarters each year.

The data employed were taken at long-cadence (LC) whereby 270 readouts of slightly more than $6.5$ second duration ($6.01982$ second integration and $0.51895$ second read time) are co-added to $29.4$-minute intervals.  Quarters 1 through 6 yield flux time-series with 1,639, 4,354, 4,370, 4,397, 4,633, and 4,397 cadences (see also Table~\ref{tab:observations}) corresponding to $33.5$, $88.9$, $89.3$, $90.3$, $94.7$ and $89.8$ days of photometry, respectively.  The exception to this is the number of cadences in Quarter 4 for targets falling on CCD module 3 (channels 5, 6, 7, and 8).  Such targets were observed for 1,022 cadences instead of 4,397.  Besides the interruption for some targets due to the Module 3 failure, each quarterly time-series contains gaps, some larger than others, due to a variety of occurrences including monthly breaks for data downlink, occasional safe mode events, manually excluded cadences, loss of fine point, and attitude tweaks.  All missing cadences are tabulated in the Anomaly Summary Table in Section 5 of the Data Release Notes (DRN) archived at MAST\footnotetext{{\it Multi-Mission Archive at Space Telescope Science Institute}; \url{http://archive.stsci.edu/kepler}}.  Also included in the DRN are the start and stop time of each quarter.  This information, together with the transit ephemerides presented in Table~\ref{tab:planetProps1} is sufficient for reconstructing the number of observed transits in time-series of any length.

Pixel data are converted to instrumental fluxes via \ek\ pipeline software modules that calibrate pixel data \citep{quintana}, perform aperture photometry \citep{twickena}, and correct for systematic errors \citep{twickenb}.  The pipeline software is documented in the {\it Kepler Data Processing Handbook} (KSCI-19081) at MAST.  As described in Section 2 of that document, each data set is associated with a software release number (SOC version number).  For this analysis, Quarters 1 through 4 were processed with SOC 6.1 code, Quarter 5 was processed with SOC 6.2, and Quarter 6 was processed with a pre-release version of SOC 7.0.  Specifics about the features of each can be found in the DRN at MAST that accompany each release.  Note that quarterly data are reprocessed as new pipeline versions become available.   Information about the data utilized herein can be found in Data Release Notes 4 through 9. We note that the Quarter 6 data archived at MAST may differ slightly from the data employed here since the latter were processed with pre-release software. 

\section{Transit Identification}
\label{sec:tce}

Systematic-error corrected light curves for all quarters under consideration are passed to the Transiting Planet Search (TPS) pipeline module to identify signatures of transiting planets. The functionality of this module is described in  \cite{jenkinsSPIE}, \cite{tenenbaum}, and in the {\it Kepler Data Processing Handbook} at MAST. Before searching for transits, the software stitches together each quarterly data segment to form one contiguous light curve. To accomplish this, TPS removes a polynomial fit constrained to achieve zero offset and zero slope in the first and last day of each quarter and to ensure that the result is approximately zero-mean and wide sense stationary. TPS then identifies and removes strong sinusoidal features from each quarterÕs light curve via  a periodogram-based approach. Finally, the gaps between quarters are filled via an autoregressive modeling technique to condition the time-series for the FFT-based detection algorithm.

Transit signals are identified using an adaptive, wavelet-based matched filter that explicitly takes the Power Spectral Density (PSD) of the observation noise (stellar variability + shot noise + residual instrument noise)  into account in formulating the detection statistics for each light curve. TPS transforms the time-series into the wavelet domain (a joint-time/frequency analysis) in order to characterize the power spectral density as a function of time and then correlates a transit pulse of a given duration  with the normalized, conditioned light curve in this wavelet domain to generate a correlation statistic time-series that measures the likelihood that a transit is present at each time step. The single event correlation time-series is then folded over each trial period from the minimum (0.5~days) to the length of the data set to identify Threshold Crossing Events (TCEs): instances where a given period and epoch exceed the detection threshold of $7.1\sigma$. Fourteen distinct transit searches are conducted, each using a different pulse duration (1.5, 2.0, 2.5, 3.0, 3.5, 4.5, 5.0, 6.0, 7.5, 9.0, 10.5, 12.0, 12.5, and 15 hours). The maximum multiple event statistic (MES) over all durations is an estimate of the ratio of the  transit depth to the uncertainty in the transit depth as a fitted parameter for the given rectangular transit pulse train at which the maximum occurred. At a threshold of $7.1\sigma$, fewer than one false alarm is expected over the baseline mission duration due to statistical fluctuations. This defines the threshold for consideration as a viable candidate. Often, several of the matched filter pulse durations  yield a detection statistic that passes our criterion (as expected), in which case the highest value is adopted. MES values are presented in column 14 of Table~\ref{tab:planetProps2}.  291 candidates are assigned MES values of -99 signifying an invalid value. This occurs when the period returned by the pipeline is not the final period derived from full light curve modeling.\footnote{TPS usually gets the correct period for single transit signatures but occasionally chooses a multiple of the true period as the maximum, and is sometimes confused by light curves with multiple transiting planet signatures.} MES values of -99 also occur for the new ``multis'' (additional candidates associated with stars already having at least one candidates) identified by non-pipeline products (see below).

The analysis presented here stems from a TPS run used for verification and validation (V\&V) of the pre-release SOC 7.0 pipeline.  It was the first time that TPS was run using the multi-quarter functionality.   Such functionality was not available for the production of the B11 catalog.  Long period candidates in the B11 catalog were identified using a modified Box Least Squares (BLS) algorithm (see, for example, \citealt{bls}).  

TPS returned 104,999 unique targets with at least one sequence of periodic transits yielding MES $>7.1$ (i.e. TCEs).  The majority of these TCEs are triggered by transients in the normalized light curves.  Hence, the list is further culled by applying a second criterion based on the ratio of the multiple event statistic (MES) to the maximum single event statistic (SES) contributing to the MES.  The single event statistic is the maximum correlation statistic in the time domain at a given test period.  MES/SES should be comparable to the square-root of the number of observed transits.  \cite{tenenbaum} shows that there are two distinct populations of TCEs, with a dividing line at MES/SES $=\sqrt{2}$.  Below this value, detections are likely the result of two highly unequal single events as opposed to two legitimate transits of equal depth and duration.  Discarding TCEs with MES/SES $<\sqrt{2}$ reduces the number of viable TCEs from 104,999 to 4,531. 

A transit model \citep{man02} is applied to all 4,531 remaining TCEs, and the results are manually inspected to identify obvious false alarms and other astrophysically interesting signals that are clearly not consistent with the planet interpretation.  The result is a list of 1,058 stars that were then assigned \ek\ Object of Interest (KOI) numbers.  All 1,058 stars, as well as those reported in the B11 catalog, were searched for evidence of additional transit sequences.   The search was performed by subtracting the transit model for the primary TCE, and then passing the residual to the modified BLS transit detection software. This yielded an additional list of 332 candidates associated with known KOIs.  As in \cite{febCatalog}, decimal values (.01, .02, .03, ...) are added to the KOI number to distinguish between multiple candidates associated with the same star.  They  are assigned in the order they were identified.

\section{Candidate Vetting}
\label{sec:vetting}

The procedures described in Section~\ref{sec:tce} produce 1,390 KOIs that are then vetted for astrophysical false positives in the form of eclipsing stellar systems.  Here, we describe the suite of statistical tests employed.  They are separated by the type of data they operate on.  Statistical tests derived from the flux time-series and the corresponding transit models are described in Section~\ref{sec:dv}, while statistical tests derived from pixel-level data are described in Section~\ref{sec:photocenter}.  Both make use of pipeline products as well as off-line analyses.  The pipeline module that evaluates KOIs for likely false positives, referred to as Data Validation (DV), is described by \cite{dv} and the {\it Kepler Processing Handbook} at MAST.  Section~\ref{sec:procedures} describes the overall procedures that were followed to promote a KOI to planet candidate status.

As work to identify and vet candidates progressed, new products became available.  Quarter 7 and Quarter 8 photometry, for example, was available in the summer of 2011.  These data were utilized in the off-line (i.e. non-pipeline) light curve modeling used to determine various vetting metrics, as well as the properties of the the planet candidates tabulated in Tables~\ref{tab:planetProps1} and~\ref{tab:planetProps2} and described Section~\ref{sec:modeling}.  Moreover, testing of pre-release SOC 8.0 code (also using Q1-Q8 data) in the fall of 2011 produced significantly improved Data Validation reports and metrics and were also used to vet the Q1-Q6 candidates reported in this contribution.

A subset of the vetting metrics used for candidate evaluation are provided in Table~\ref{tab:planetProps2} so that users can identify the weaker candidates and know what types of problems to look for.  These metrics are described in turn below and summarized in Section~\ref{sec:procedures}.

\subsection{Tests on the Flux Time-series}
\label{sec:dv}

For each KOI, the even-numbered transits and odd-numbered transits are modeled independently using the techniques described in Section~\ref{sec:tce}.  The depth of the phase-folded, even-numbered transits is compared to that of the odd-numbered transits as described in \cite{batalha_fp} and \cite{dv}.  A statistically significant difference in the transit depths is an indication of a diluted or grazing eclipsing binary system.  A similar metric is computed by the DV pipeline module as described by \cite{dv}.  Each uses a different methodology for detrending the light curves (i.e. filtering out stellar variability), and both proved useful and are tabulated in columns 6 (modeling-derived statistic: $O/E_{\rm 1}$) and 7 (DV-derived statistic: $O/E_{\rm 2}$) of Table~\ref{tab:planetProps2}.  In general, 3-$\sigma$ was the threshold for flagging a KOI as a false positive.  However, when the two values disagreed,  we deferred to the Data Validation statistic owing to its more sophisticated whitening filters.  143 KOIs in Table~\ref{tab:planetProps2} have $O/E_{\rm 1} > 3$ while only 7 have $O/E_{\rm 2} > 3$.  Five have both $O/E_{\rm1}$ and $O/E_{\rm 2}$ larger than 3-$\sigma$ but are otherwise clean candidates.  Further inspection of their light curves suggested that stellar variability and/or instrumental transients were driving an anomalously high odd/even statistic, and the candidates were retained.  In 121 cases, the DV model fitter failed, thereby precluding quantification of an odd/even statistic.  In 18 of these cases, $O/E_{\rm}1$ is larger than 3-$\sigma$ and the candidate was retained anyway.   While most of these are marginal cases near the 3-$\sigma$ cutoff, users are cautioned that exceptions exist and should be examined independently on a case-by-case basis.

The modeling allows for the presence of a secondary eclipse (or occultation event) near phase=0.5 as a means of identifying diluted or grazing eclipsing star systems. The Secondary Statistic (column 8 of Table~\ref{tab:planetProps2}) is the relative flux level at phase 0.5, divided by the noise.  As such, it can have positive as well as negative values.  While its presence does not rule out the planetary interpretation, it acts as a flag for further investigation.  More specifically, the flux decrease is translated into a surface temperature assuming a thermally radiating disk, and this temperature is compared to the equilibrium temperature of a low albedo (0.1) planet at the modeled distance from the parent star.  If the flux change is not severe enough to rule out the planetary interpretation (ascertained by the difference between the surface temperature and equilibrium temperature), the candidate is retained.   Eight KOIs retained in the catalog have a statistic outside of (negative) 3$\sigma$.  In each of these cases, it appears possible that the occultation signal is a result of stellar and/or instrumental flux changes.  This statistic is relevant primarily for short-period orbits where circularization is expected since the search is only done at phase 0.5.  The Data Validation pipeline module checks to see if additional transit sequences were identified in the light curve at the same period but different phase.  No such sequences were identified for the candidates reported here.

Table~\ref{tab:vshaped} lists KOIs (both old and new) that are V-shaped.  The shape of a transit is not used as a diagnostic for rejecting planet candidates.  The right combination of properties and geometry can, indeed, produce ingress and egress times that are a significant fraction of the total transit duration (e.g. grazing transits).  However, a diluted eclipsing binary system is another possible interpretation and does not require such a narrow range of inclination angles (impact parameters).  The false positive rate amongst the V-shaped candidates is expected to be higher than the false positive rate of the general population.  A metric is constructed to flag such cases: $1-b-\frac{\rpl}{\rstar}$.  Negative values imply that the purported planet is not fully covering the stellar disk at mid-transit.  The closer the number is to $-2\times \frac{\rpl}{\rstar}$, the more severely it is grazing.  A grazing geometry is required to model V-shaped transits that are not caused by a large planet-to-star size ratio.  The KOIs with light curves modeled as grazing transits are listed in Table~\ref{tab:vshaped} together with the orbital period and reduced radius.  

\subsection{Tests on the Pixel Data}
\label{sec:photocenter}

A major source of false-positive planetary candidates in the \ek\ data is a background eclipsing binary (BGEB) star within the photometric aperture of a \ek\ target star.  These BGEBs, when diluted by a target star, can produce mimic a planetary transit signal.  Two methods are used to detect such BGEBs by using the \ek\ pixels to determine the location of the object causing the transit signal: direct measurement of the source location via difference image analysis and inference of the source location from photo-center motion associated with the transits.

Photo-center motion is measured by computing the flux-weighted centroid of all pixels downlinked for a given star both in and out of transit (the times of which differ for each candidate associated with a given star).  The computation of the flux centroids for KOIs that were processed with the SOC pipeline Data Validation module is done per cadence, and the centroid offset when in transit is computed as the amplitude of a joint multi-quarter least-squares fit to the transit model generated by the SOC pipeline \citep{dv}. For those targets that were not processed by DV, the flux centroid offset is computed for the quarterly average out-of-transit and in-transit pixel images.  In this latter case, the centroid offsets are computed by subtracting the out-of-transit centroid from the in-transit centroid, and the quarterly offsets are averaged as described below.  Once the in-transit centroid offset is computed, the source location is then inferred by scaling the difference of these two centroids by the inverse of the flux as described in \cite{kepler8b}.  

Difference image analysis takes the difference between average in-transit pixel images and average out-of-transit images.  Barring pixel-level systematics and field-star variability, the pixels with the highest flux in the difference image form a star image at the location of the transiting object, with flux level equal to the fractional depth of the transit times the original flux of the star. Performing a fit of the \ek\ pixel response function (PRF) \citep{Bryson:10} to both the average difference and out-of-transit images gives the sky location of the transit source and the target star.  The offset of the transit source from the target star is then defined as the transit source minus target star location.  For most KOIs, the difference image offset is computed per quarter, and the quarterly offsets are averaged as described below.  For a small number of KOIs with very low SNR, a computationally expensive joint multi-quarter fit is performed, with uncertainties estimated via a bootstrap analysis.

In principle, both the photo-center motion and difference image techniques are similarly accurate for isolated stars and sufficiently high SNR transits, but the techniques have different responses to systematic error sources such as field crowding.  The photo-center method is more sensitive to noise for low SNR transits and crowding by field stars.  In particular, the estimate of the transit source location by scaling the offset is highly sensitive to incompletely captured flux for either the target star or field stars in the aperture.  

In the difference image method, the PRF fit to the difference and out-of-transit pixel images is biased by PRF errors described in \cite{Bryson:10}, as well as errors due to crowding.  Defining the centroid offset as the difference between the out-of-transit and difference images nearly cancels the PRF bias because both fits are subject to the same error.  Bias due to crowding, however, is more of an issue.  Excepting cases in which background stars are strongly varying, a difference image removes most point sources, leaving the transit signal as the predominant change in flux.  The average out-of-transit image, on the other hand, contains all point sources.  Consequently, the offset (formed by comparing a direct image and a difference image) may contain a bias.

\begin{figure}
\begin{center}
\includegraphics[width=175mm]{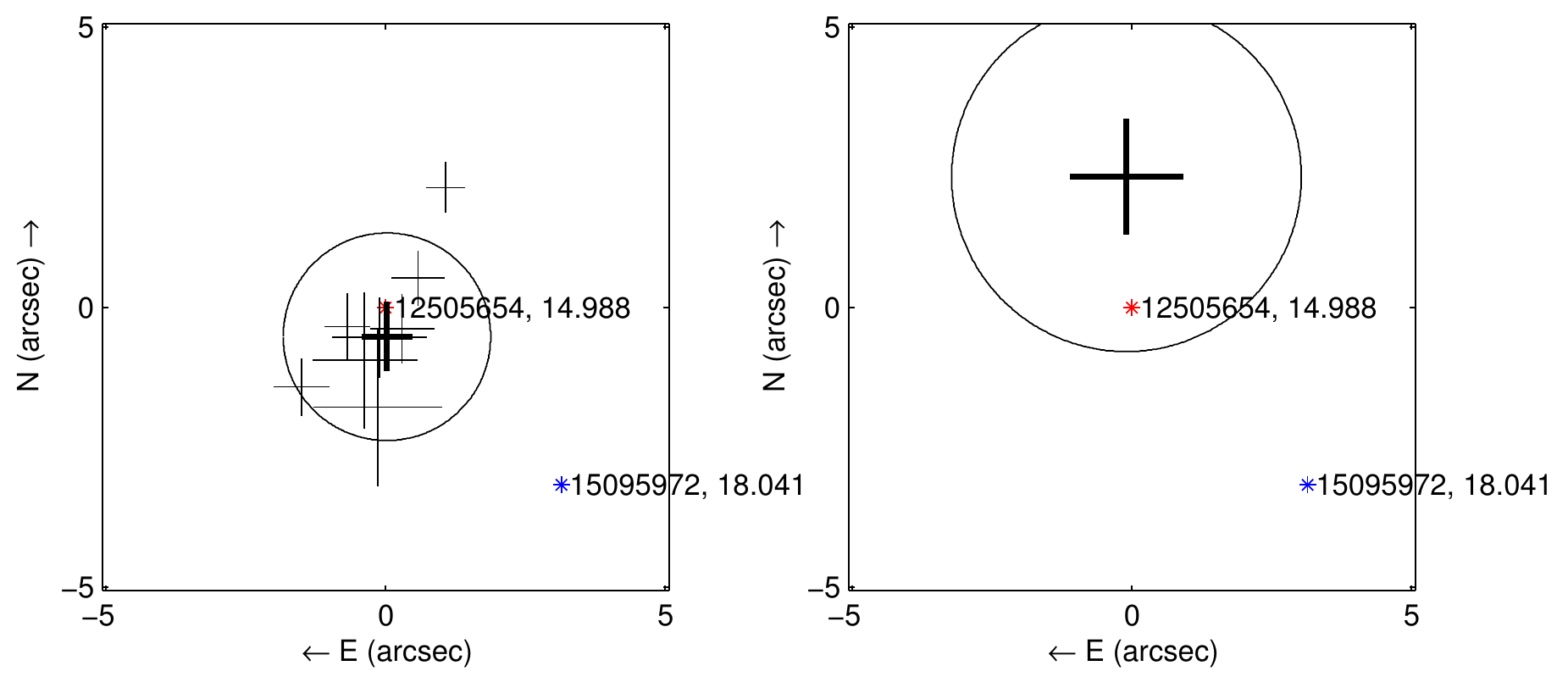}
\end{center}
\caption{Transit source location determined by the difference image (left panel) and photo-center motion (right panel) techniques for KOI-2353.  On the left, the thin crosses show the individual quarter measurements, and the bold cross shows the multi-quarter average.  On the right, the transit location inferred from the multi-quarter joint fit to the photo-center motion is shown. In both panels the length of the cross arms show the 1-$\sigma$ uncertainties in RA (x-axis) and Dec (y-axis).  The circle shows the 3-$\sigma$ uncertainty in the offset distance.  The asterisks show star locations, labeled by Kepler ID and Kepler magnitude, with the centered asterisk showing the location of the target star.  This is a clear example where the transits are associated with the target, rather than the nearby, faint background star.}
\label{fig:diffimage}
\end{figure}

Both bias types vary from quarter to quarter. A study of a large number of targets indicates that the combined biases have an approximately zero-mean distribution across quarters, so we can reduce their impact by averaging the quarterly centroid measurements across quarters.  We compute this average by performing a $0^{\rm th}$ order (constant) polynomial robust $\chi^2$ minimizing fit to the quarterly results.  This approach produces an uncertainty that takes into account both the uncertainty of the individual quarterly observations as well as their scatter due to bias across quarters.  An example set of quarterly measurements and the resulting average is shown in the left panel of Figure~\ref{fig:diffimage}.  This method works best for short-period candidates, where there are many transits in all quarters.  This method is less effective in reducing the centroid bias for long-period candidates. In particular, single-transit candidates that appear in only one quarter may have unknown biases in their centroid positions that are not accounted for in the centroid uncertainties.

The above centroid analysis was performed using data from Quarters 1 through 8.  The transit source location offsets reported in Table~\ref{tab:planetProps2} are from the difference image method since it is more reliable as evidenced by consistently smaller uncertainties compared to the photo-center motion method.  A candidate passes the photo-center vetting step when its multi-quarter transit source offset is less than $3\sigma$.  There are, however, exceptions. KOIs having transit source offsets larger than $3\sigma$ were identified as having systematic errors (e.g. crowding biases) that are likely causing the large apparent offset.  Modeling efforts to confirm these systematics are currently underway.  

In some cases, the PRF fit algorithm failed, typically due to low quarterly SNR or to bright field stars that prevented reliable determination of the target star location via PRF fit.  We retain these targets when visual inspection of the difference image indicates that the change in flux due to the transit is on the target star.  Finally, difference imaging is very inaccurate for saturated targets, and we retain saturated candidates for which the difference images show no obvious indication of a background source.  Offset values for slightly saturated stars (Kp between 10.5 and 11.5) are likely accurate to within $4\arcsec$ (one pixel), while offset values for highly saturated stars (Kp $< 10.5$) should be disregarded.  Transit source offset values are set to -99 when we feel that the centroid measurement is unreliable.

\subsection{Promotion to Planet Candidate}
\label{sec:procedures}

KOIs were divided amongst more than twenty science team members for evaluation of the following metrics:  1) odd/even statistic, 2) occultation test,  3) quality of model fit, 4) long/short period comparison, 5) single-quarter photo-center motion, and 6) multi-quarter photo-center motion.  An integer value of 0, 1, or 2 was assigned to each metric to indicate if the test clearly passed (0), was ambiguous (1), or clearly failed (2).  A similar flag was ascribed to the visual appearance, where examples of suspicious characteristics would include markedly V-shaped transits, red noise and/or outliers in the time-series calling into question the reliability of the transit signal, anomalously long transit durations, poor light curve fits, obvious secondaries, etc.   Each candidate was then designated ``yes'', ``no'', or ``maybe''.  Candidates flagged as ambiguous (maybe) were re-evaluated.  In most cases, this required updated light-curve modeling, more detailed inspection of the software pipeline products, and/or further scrutiny of the pixel flux analysis yielding photo-center statistics.  

Once every KOI was assigned a ``yes'' or ``no'' designation, the integer flags were summed and the distribution for the two populations was compared.  This elucidated a small number ($<10$) of inconsistent assignments. 

The period and location on the sky of every new KOI was cross-checked against the list of previously known KOIs and the catalog of Eclipsing Binaries \citep{eb1,eb2}.  This is a safeguard against redundancy.  More importantly, though, the cross-check serves to identify flux contamination from bright Eclipsing Binaries that the photo-center analysis missed.  Approximately 20 targets within $20\arcsec$ of an existing KOI or EB with the same orbital period (to within 5 minutes) were identified.  In this manner, we also identified two small swarms of spatially co-located stars, each with a transit-like event at the same period of its neighbors.  The lack of sizable photo-center motion and the spatial extent of the contaminating flux suggests that scattered light (e.g. an optical ghost from a bright Eclipsing Binary at the focal plane anti-podal position) is responsible.

Every one of the 1\,390 KOIs evaluated here will appear either in this contribution as a viable planet candidates, or in the associated false positive catalog \citep{brysonFP}, or in future versions of the Eclipsing Binary Catalog. 

Eight high-SNR ($>30\sigma$) single transit events were identified and included in the catalog.  Their corresponding orbital periods are estimated from the transit duration and knowledge of the stellar radius assuming zero eccentricity.  The periods are then rounded to the nearest integer and multiplied by $-1$ so as to distinguish them from the candidates that have reliable orbital ephemerides.  Parameters requiring an accurate period (e.g., odd/even statistic, semi-major axis, etc) are set to $-99$, which is the value adopted globally for unreliable, spurious, or unknown values.

The final list of viable planet candidates is presented in Table~\ref{tab:planetProps1} with additional information listed in Table~\ref{tab:planetProps2}.  In addition to the Multiple Event Statistic (MES) from the Q1-Q6 TPS run, the following vetting metrics are provided:  a) the odd/even statistic that tests for an eclipsing star system of nearly equal-mass components at twice the period (columns 6 and 7), b) the occultation (secondary) statistic that tests for a weak secondary event inconsistent with a planet occultation at phase 0.5 (column 8), c) the photo-center offsets in RA and DEC measured in arcsec and their associated uncertainties (columns 9 through 12), and d) the total photo-center offset position measured in units of the noise (column 13).  Analysis is based on a blend of both quantitative metrics and manual inspection.  Both the promotion from TCEs to KOIs and the promotion of KOIs to planet candidates has a human element that not only increases the reliability of the catalog but also reduces the number of false negatives that are discarded.   

\section{Properties of Planet Candidates}
\label{sec:planetProps}

Here, we describe the light curve modeling that characterizes both the orbital and physical characteristics of each planet candidate.  

\subsection{Model Fitting}
\label{sec:modeling}

For each KOI, a transit model was fit to the data.  The transit model uses the analytic formulae of \citet{man02} to model the transit and a Keplerian orbit to model the orbital phase.  The model fits for the mean stellar density (\rhostar), center of transit ($T_0$), orbital period ($P$), scaled planetary radius ($\rpl/\rstar$), impact parameter ($b$) and occultation (secondary) depth (at phase=0.5).   In the case of multiple transiting candidates the orbits were assumed to be non-interacting and $T_0$, $P$, $b$ and the occultation depth were fit for each candidate.  The assumption is made that all candidates in the same system orbit the same star modeled by \rhostar\ and that the total mass of the companion is much less than the mass of the central star.  For a Jupiter-mass companion, an error of 0.02\% will be incurred on the measurement of \rhostar, a $0.1~M_\sun$ companion would skew our estimate of \rhostar\ by approximately $2\%$, and a 0.5~$M_\sun$ companion would induce a systematic error of 41\% on \rhostar.  These assumptions give:
\begin{equation}\label{eq:rhostar}
\left(\frac{a}{\rstar}\right)^3 \frac{\pi}{3 G P^2} = \frac{(M_\star+M_{\rm p})}{\frac{4 \pi}{3}R_\star^3} \approx \frac{M_\star}{
\frac{4 \pi}{3}R_\star^3}=\rhostar,
\end{equation}
where $a/\rstar$ is the reduced semi-major axis.  The orbits were assumed to be circular, thus the transit is only sensitive to the planet-star separation for a small portion of the orbit and in general is not a true determination of \rhostar\ or $a/\rstar$.  As such, we purposely report $d/\rstar$ which is the ratio of the planet-star separation during transit to the stellar radius.  It is equivalent to $a/\rstar$ for the case of zero eccentricity.  A full orbital solution is required to correctly determine stellar parameters from transit modeling.  Great care must be taken when interpreting the fitted stellar parameters.  A mismatch between the reported stellar parameters and a transit-derived stellar parameter can be interpreted as evidence of: an eccentric orbit, erroneous stellar classification or a transiting companion with significant mass.  One is not able to distinguish among these three cases using the model parameters provided. Best fit model parameters were determined by computing a least-squares chi-squared search using a Levenberg-Marquardt algorithm \citep{pre92}.  Derivatives were numerically determined.  Uncertainties are taken from the diagonal elements of the formal covariance matrix corresponding to each fitted parameter.  They do not account for correlated errors.

\subsection{Stellar Parameters}
\label{sec:starProps} 

Output parameters from light curve modeling can be used to compute the radius, semi-major axis, and equilibrium temperature of each planet candidate given knowledge of the stellar surface temperature, radius, and mass.  These and other stellar properties for the host stars are listed in Table~\ref{tab:starProps}.  Stellar coordinates and the apparent magnitude in the \ek\ bandpass are obtained from the \ek\ Input Catalog (KIC) at MAST \citep{kic}.  

\begin{figure}
\begin{center}
\includegraphics[width=175mm]{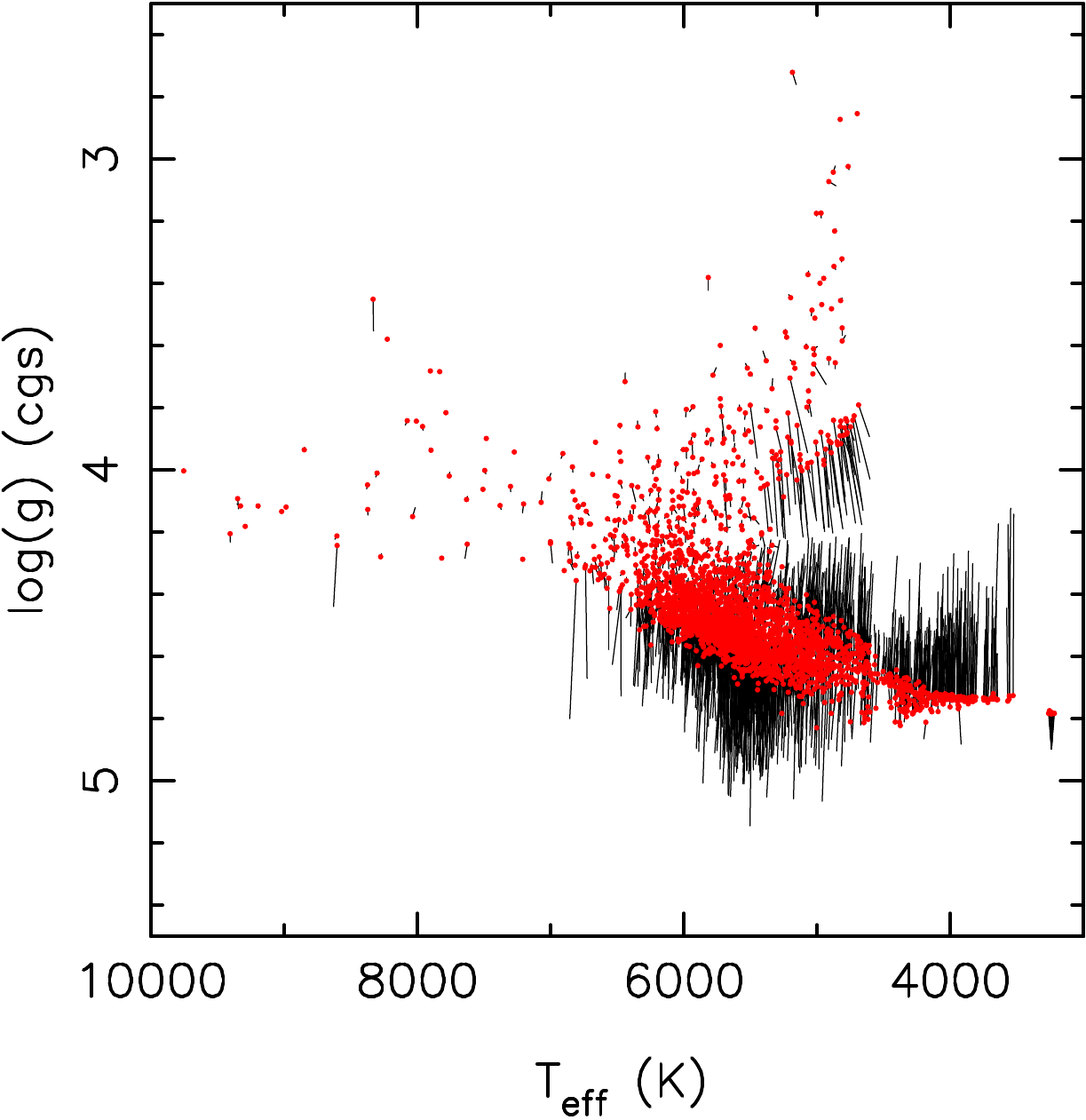}
\end{center}
\caption{Surface gravity versus effective temperature for the host stars.  The red dots correspond to updated values based on a search of the Yonsei-Yale evolution models.  Black lines point back to the locations defined by the KIC values.}
\label{fig:KICupdate}
\end{figure}

The \ek\ Input Catalog has been a valuable resource for the stellar properties necessary for characterizing {\it Kepler's} planet candidates (e.g. \teff, \logg, and \rstar).  However, there are parameters in the KIC that imply populations that are inconsistent with stellar theory and observation.  For example, G-type stars with $\logg\ \approx 5$ (i.e. well below the main-sequence) are not uncommon in the KIC.  While the uncertainty on the determination of \logg\ shows that such a star is consistent with being near the main-sequence, using such stellar parameters results in small stellar and planetary radii and an underestimation of incident flux on the planet candidate.  We systematically correct such populations in the KIC by matching \teff, \logg\ and \feh\ to the Yonsei-Yale stellar evolution models \citep{yonsei}.  We find the closest match based on minimization of:
\begin{equation}\label{eq:chisq}
\chi^2 = \left(\frac{\delta \teff}{\sigma_{\teff}}\right)^2 + \left(\frac{\delta \logg}{\sigma_{\logg}}\right)^2 + \left(\frac{\delta \rm \feh}{\sigma_{\rm \feh}}\right)^2,
\end{equation}
where $\delta$ represents the difference in the KIC and Yonsei-Yale parameter and $\sigma$ is the adopted uncertainty in the KIC parameter.  We adopted $\sigma_{\teff}=200$~K, $\sigma_{\logg}=0.3$ and $\sigma_{\rm \feh}=0.4$ and required the modeled age to be less than 14 Gyr.  For each star we adopt the model-determined mass and radius.  Figure~\ref{fig:KICupdate} shows \teff\ and \logg.  The red dots show updated values that have been adopted for determination of the stellar mass and radius.  Stellar properties derived in this manner are flagged in Table~\ref{tab:starProps} by setting $f_{\rm \teff}$ equal to 1.  The black lines point to the star's original location based on the KIC.  Most of the motion is in the \logg\ co-ordinate as it is always possible to find a match to \teff\ and \feh\ in the model suite. 

There are three populations that show significant changes to the determined \logg.  (1) Stars that fall below the main-sequence move to smaller values of \logg.  These stars have \teff\ from roughly $4500$ K to $6500$ K.  In general, estimates of the properties of these stars become larger and more luminous, which reduces the number of small stars and increases the amount of incident flux on the orbiting companion.  (2) Most stars cooler than $4500$~K see a substantial decrease in \logg.  However, modeled masses and radii are highly uncertain in this temperature range and should be used with caution.  (3) There is a population of stars which have KIC \logg\ near 4.1 and \teff\ near $5000$~K.  Stars in this region have no match when we restrict model ages to $\le14$~Gyr.  Such stars either move towards the RGB (lower values of \logg) or towards the sub-giant branch (larger values of \logg).  

Forty-nine host stars listed in Table~\ref{tab:starProps} have revised stellar properties determined from high-resolution spectroscopy taken as part of the \ek\ ground-based Follow-up Observation Program using the Shane 3-m telescope at Lick Observatory, the 1.5-m Tillinghast reflector at Whipple Observatory, the  2.7-m Harlan Smith telescope at McDonald, the 2.5-m Nordic Optical Telescope at La Palma, Spain, and the 10-m Keck telescope using HIRES.  Fourteen of these were subjected to LTE spectral synthesis using SME \citep{valenti96, valenti05}.  The resulting surface gravity, effective temperature, and metallicity are used to identify the best-matching Yonsei-Yale stellar evolution model \citep{yonsei} as described above for the KIC parameter adjustments. Stellar properties derived in this manner are flagged in Table~\ref{tab:starProps} by setting f$_{\rm \teff}$ equal to 3.

The spectra of thirty-five host stars were compared to a synthetic library of spectra as part of the Stellar Parameter Classification (SPC) effort and analysis tool described by \cite{metallicity}.  As in the case of the SME analysis, the resulting stellar parameters are compared to the Yonsei-Yale models to determine the stellar radius.  The revised stellar properties are listed in Table~\ref{tab:starProps} and flagged with $f_{\rm \teff}$ equal to 2.  

Twenty-one host stars have neither KIC classifications nor spectroscopically-derived stellar properties.  In these cases ($f_{\rm \teff}$ equal to 0), the effective temperature is estimated from linear interpolation of the Main Sequence properties of \cite{schmidt-kaler} at the KIC J-K color.  There is little justification for the assumption of a Main Sequence luminosity class.  False positives and large errors in the planet candidate radius should be expected amongst this sample.

Also included in Table~\ref{tab:starProps} is the rms Combined Differential Photometric Precision (CDPP) -- a measure of the photometric noise (including stellar and instrumental sources) on a 6-hour timescale after systematic-error correction and removal of strong sinusoidal features.  A detailed description of the CDPP can be found in \cite{jenkinsSPIE}.  CDPP values are crucial for statistical analyses of planet occurrence rates as they define the observational detection sensitivities.  3, 6, and 12-hour rms CDPP values for all observed stars are archived at MAST and can be obtained using the Data Retrieval Search form.

\subsection{Derived Parameters}
\label{sec:derivedParams}

The planet radius, semi-major axis, and equilibrium temperature of each planet candidate are computed using the estimated stellar properties and the parameters returned by the light curve modeling described in Section~\ref{sec:modeling}.  Planet radius (column 3 of Table~\ref{tab:planetProps2}) is the product of the reduced radius, \rpl/\rstar\ (column 11 in Table~\ref{tab:planetProps1}) and the stellar radius in column 9 of Table~\ref{tab:starProps}. Planet radii are given in units of Earth-radii.  

The semi-major axis provided in column 4 of Table~\ref{tab:planetProps2} is derived from Newton'Õs generalization of KeplerÕ's third law given the orbital period (column 2) and the stellar mass (column 10 of Table~\ref{tab:starProps}).  The stellar mass is derived directly from the surface gravity and stellar radius (column 8 and 9 of Table~\ref{tab:starProps}).    We note that, despite the name, the product of the reduced semi-major axis, $d/\rstar$ (column 9) and the stellar radius will not yield a reliable estimate of the semi-major axis of the orbit.  The reduced semi-major axis is a parameter derived from light curve modeling assuming zero eccentricity.  More specifically, it is the ratio of the star-planet separation {\it at the time of transit} to the stellar radius.  Only for the special case of either zero eccentricity or alignment of the periastron with the line of sight is it equivalent to the ratio of the semi-major axis to the stellar radius.  Similarly, there is a direct relationship between $a/\rstar$ and the stellar density (see Equation~\ref{eq:rhostar}) as described in \cite{seager03}.  However, inferring the stellar density from $d/\rstar$ can produce unphysical results. 

The equilibrium temperature, $T_{\rm eq}$ (column 5 of Table~\ref{tab:planetProps2}) is the temperature at which the incident stellar flux balances the thermal radiation.  It is derived by assuming that the planet and star act as gray bodies in equilibrium and that the heat is evenly distributed from the day to night side of the
planet (e.g., a planet with an atmosphere or a planet with rotation period shorter than the orbital period):
\begin{equation}\label{eq:teq}
T_{\rm eq}=\teff(\rstar/2a)^{1/2} [f(1-A_{\rm B})]^{1/4}, 
\end{equation} 
where \teff\ and \rstar\ are the effective temperature and radius of the host star, the planet at distance $a$ with a Bond albedo of $A_{\rm B}$.  The factor, $f$, acts as a proxy for atmospheric thermal circulation where $f=1$ (assumed here) indicates full thermal circulation. The Bond albedo, $A_B$, is the fraction of total power incident on a body scattered back into space, which we assume to be 30\%.  

\subsection{Period Aliasing}

Section~\ref{sec:vetting} describes the vetting procedures and the many metrics used to eliminate false positives.   One of the statistics is a measure of the significance of odd/even depth differences.  Aside from identifying eclipsing binaries, the metric can also be used as a warning that a low-SNR planetary period is a factor of two too low (as for KOI-730.03 and 191.04; \citealt{architectureI}).  As an additional step outside the general procedures, we constructed a similar statistic for comparing the depths of every third or fourth transit.  In this manner, the period of KOI-1445.02 was found to be a factor of three larger than initially determined (54 days compared to 18 days as determined the the pipeline).  Table~\ref{tab:planetProps1} contains the updated period value.  In this exercise we identified a handful of other planet candidates with low SNR, primarily based on one or two transits each\footnote{ KOI-2224.02, 1858.02, and 2410.02 have ephemerides indicative of two transits in the light curve; KOI-1070.03 and 2410.01 have ephemerides dominated by one transit apiece. }, with ephemerides predicting transits that are not detected.  When only two transits are seen, it is possible that each transit belongs to different planets that each only transit once, or, alternatively, that they are the primary and secondary eclipse of an eccentric, long-period, blended eclipsing binary.  In some cases, there may be an alternative orbital period that phases the observed transits with a gap in the data take.  More observations are required to determine a unique solution.  These period aliases show that while the catalog is generally of high reliability, additional analyses on the light-curves of low-SNR transits may generate revisions. 

\section{Distributions}
\label{sec:distributions}

The metrics and procedures described above, as applied to the Q1--Q6 data, yield \nCandidatesNewTotal\ new planet candidates, representing a gain of 88\%\ over the B11 catalog.  Eight of the new candidates are single-transit events (as indicated by negative, integer period values in Tables~\ref{tab:planetProps1} and~\ref{tab:planetProps2}).  After removing the single-transit based candidates, the remaining candidates range in size from one-third the size of Earth to three times the size of Jupiter (transit depths of 20 parts-per-million to 20 parts-per-thousand) and equilibrium temperatures from 200~K to 3800~K (orbital periods of a half a day to nearly one year).   Of the new candidates, \nearth,  \nsuperearth,  \nneptune, and \njupiter\ are Earth-size ($\rpl < 1.25 \rearth$), super-Earth-size ($1.25 \rearth \le \rpl < 2 \rearth$), Neptune-size ($2 \rearth \le \rpl < 6 \rearth$), and Jupiter-size ($6 \rearth \le \rpl < 15 \rearth$), respectively.  An additional \nlarger\ candidates are included in the catalog that are larger than $15 \rearth$, a small number of which are larger than three times the size of Jupiter and unlikely to be consistent with the planet interpretation.  They are included here due to the uncertainties in the stellar radii (see Section~\ref{sec:starProps}).   Section~\ref{sec:hz} describes the subsample of the new planet candidates that are in the Habitable Zone.

\begin{figure}
\begin{center}
\includegraphics[width=175mm]{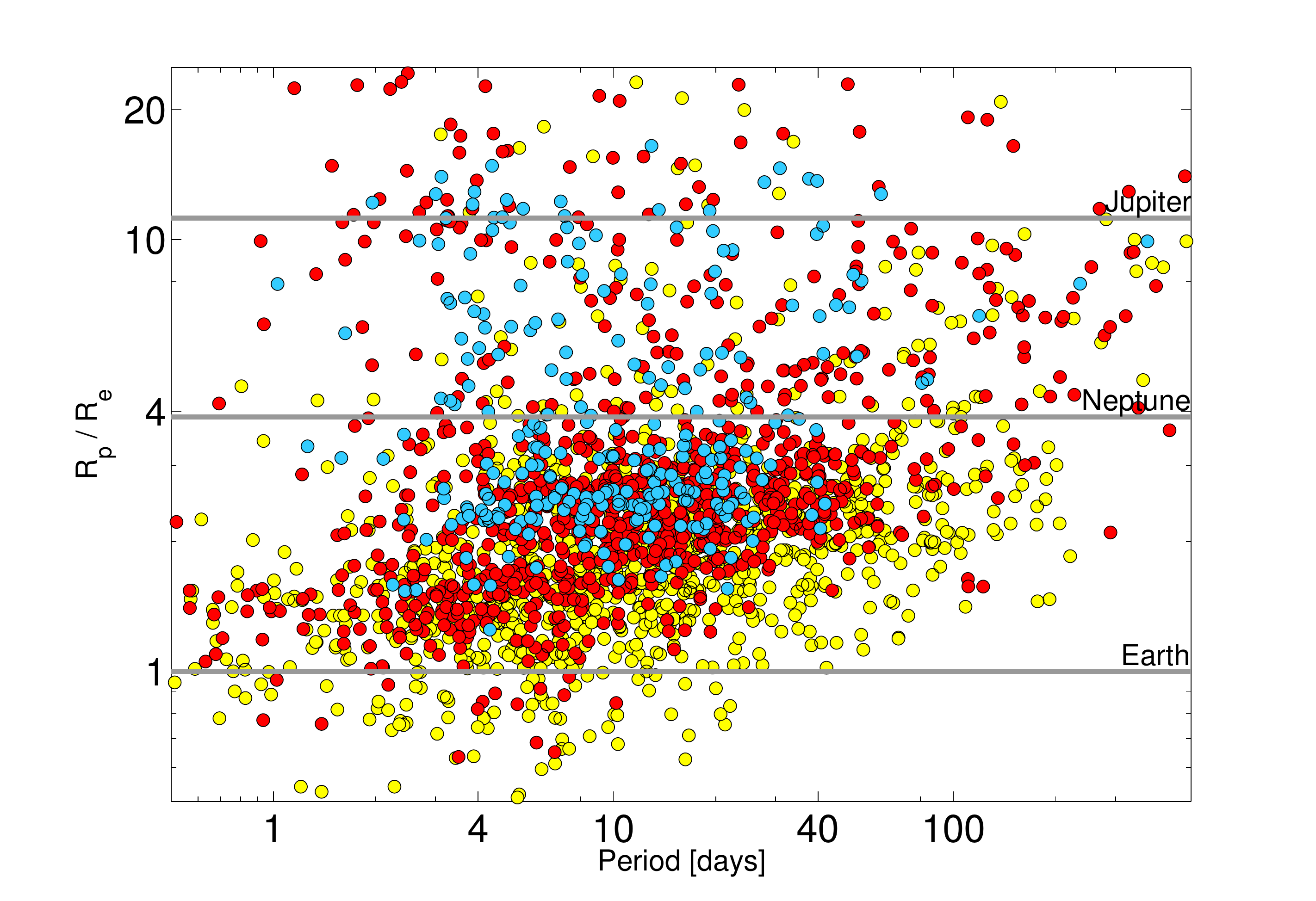}
\end{center}
\caption{Radius versus orbital period for each of the planet candidates in the B10 \citep{juneCatalog} catalog (blue points), the B11 \citep{febCatalog} catalog (red points), and this contribution (yellow points).  Horizontal lines marking the radius of Jupiter, Neptune, and Earth are included for reference. }
\label{fig:radiusPeriod}
\end{figure}

Figure~\ref{fig:radiusPeriod} shows planet radius versus orbital period for the candidates in the B10 catalog (blue) and the B11 catalog (red) together with the new candidates reported here (yellow).  The properties of all previously published KOIs have been updated as described in Appendix~\ref{sec:bigCatalog}.  The range of the abscissa and ordinate in Figure~\ref{fig:radiusPeriod} are truncated at 500~days and $25\rearth$ (to more effectively display the population) thereby excluding \nexcluded\ candidates, \nexcludedsingles\ of which display only a single transit.    The points are layered (newest candidates underneath) so that the growing domain and range of each population is apparent.  Not surprisingly, each successive catalog contains progressively smaller planet candidates at progressively longer orbital periods.  The relative number of small planet candidates is one of the striking features of the B11 catalog: over \febSmallerThanNeptune\ of the candidates presented there are smaller than Neptune.  This trend continues: in the current sample of new candidates, over \newSmallerThanNeptune\  are smaller than Neptune.

\begin{figure}
\begin{center}
\includegraphics[width=175mm]{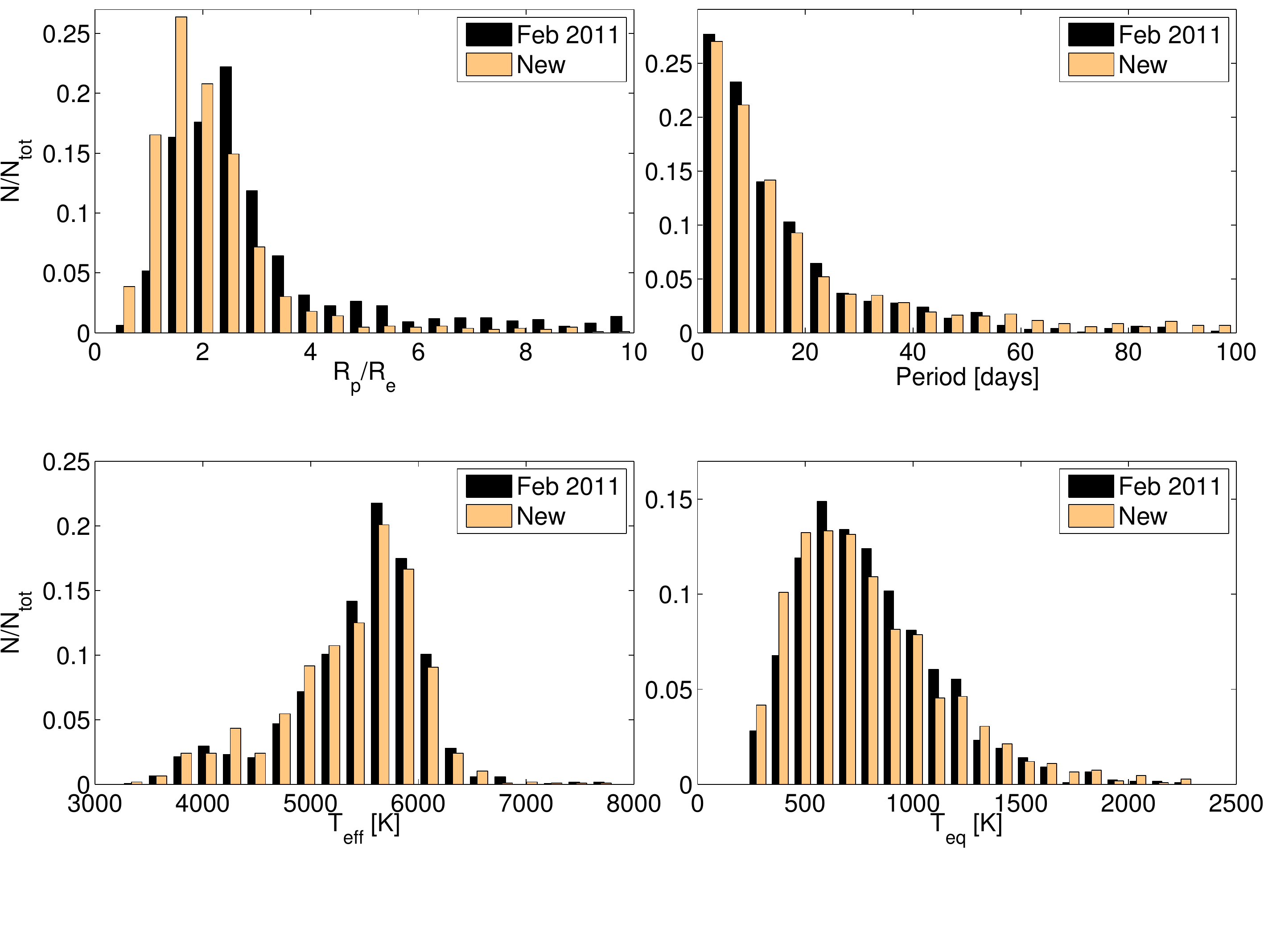}
\end{center}
\caption{Radius, Period, and $T_{\rm eq}$ distributions of the planet candidates in the B11 catalog (dark) and the new candidates presented here (light).  The distribution of the surface temperature of their host stars is also included.  Counts are expressed as fractions of the total number of candidates in each of the two catalogs.}
\label{fig:gains}
\end{figure}

The relative gains in the number of candidates are displayed in Figure~\ref{fig:gains} which contains normalized distributions of planet radius, period, equilibrium temperature and host star temperature for the B11 catalog and for the new candidates presented here. In terms of radius, the gains are predominantly for candidates smaller than $2 \rearth$.   The new candidates contain a significantly smaller fraction of Neptune-size and Jupiter-size planets than the Feb 2011 catalog.  We report a growth of \gainSmall\ for candidates smaller than $2 \rearth$ compared to \gainLarge\ for candidates larger than $2 \rearth$, and a growth of \gainLong\ for orbital period longer than 50~days compared to an \gainShort\ increase for periods shorter than 50~days.   The gains in equilibrium temperature and host star properties are more uniform  (approximately $88\%$ for most bins).  Section~\ref{sec:gains} compares these gains to what would be expected from increasing the baseline observation window from thirteen months (Q1--Q5) to sixteen months (Q1--Q6).

\section{Discussion}
\label{sec:discussion}

\subsection{Computed versus Observed Growth in Numbers of Candidates}
\label{sec:gains}

\begin{figure}
\begin{center}
\includegraphics[width=175mm]{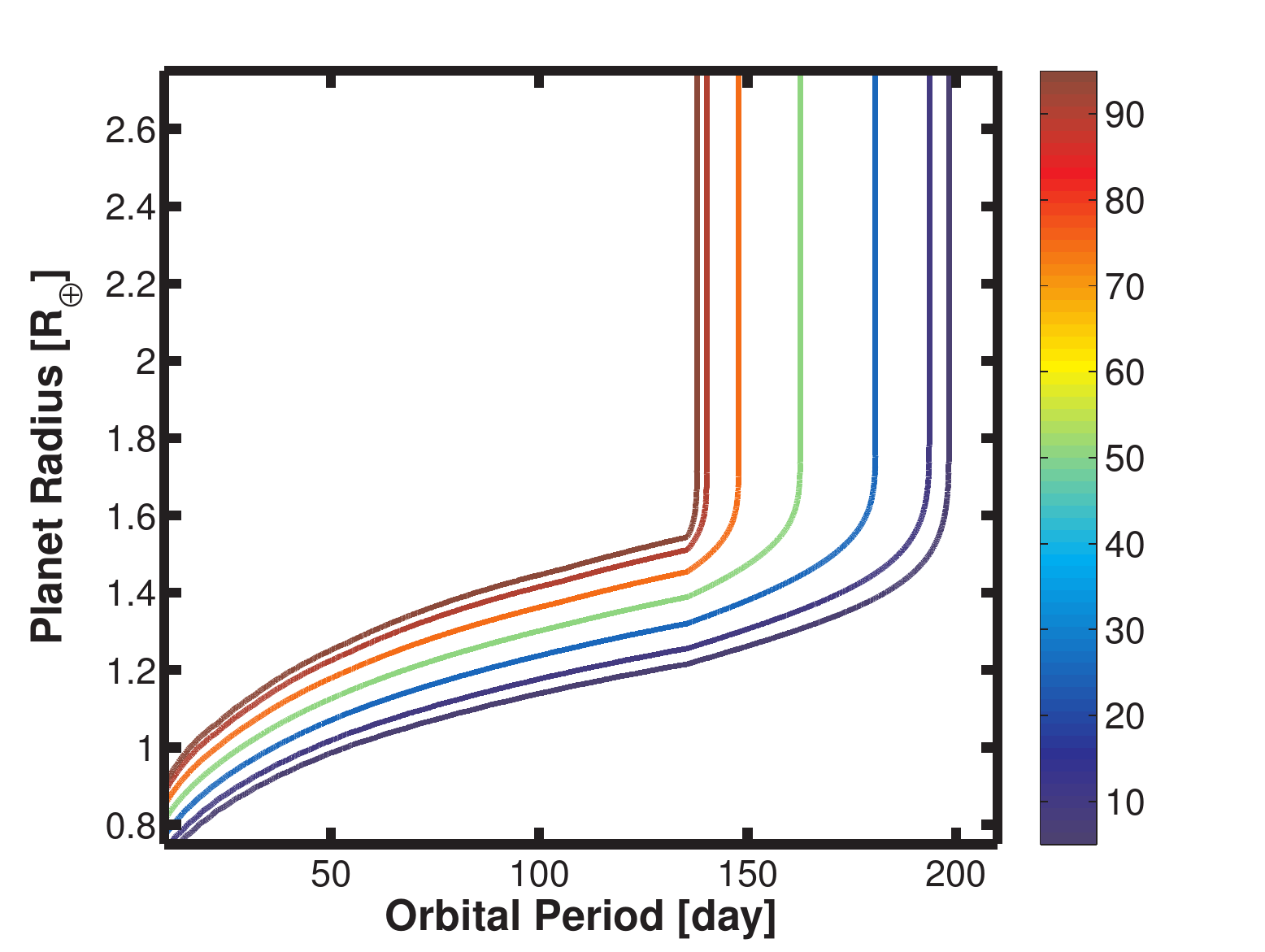}
\end{center}
\caption{Detection probability (expressed as percent completeness) contours as a function of planet radius and orbital period for a Kp=12 magnitude star with the properties listed in Table~\ref{tab:completeness1}.}
\label{fig:completeness1}
\end{figure}

We compare the observed gain in the numbers of candidates to that expected from an increase in sensitivity afforded by three additional months of data.  To do so, we express the total SNR of a transit event of a given depth, duration, and period associated with a star of specified characteristics as a detection probability.  We construct a grid of such probabilities for our hypothetical star as a function of transit depth and orbital period.  There is a unique transit duration associated with each point in the grid, computed by assuming a zero-eccentricity Keplerian orbit.  Knowledge of the duration is necessary for establishing the total SNR.  Figure~\ref{fig:completeness1} shows a sample grid for a typical Kp=12 magnitude star assuming five quarters of observations (407 days) and a 95\%\ duty cycle.  The grid is displayed as contours of equal detection probability, or ``percent completeness''.

\begin{figure}
\begin{center}
\includegraphics[width=175mm]{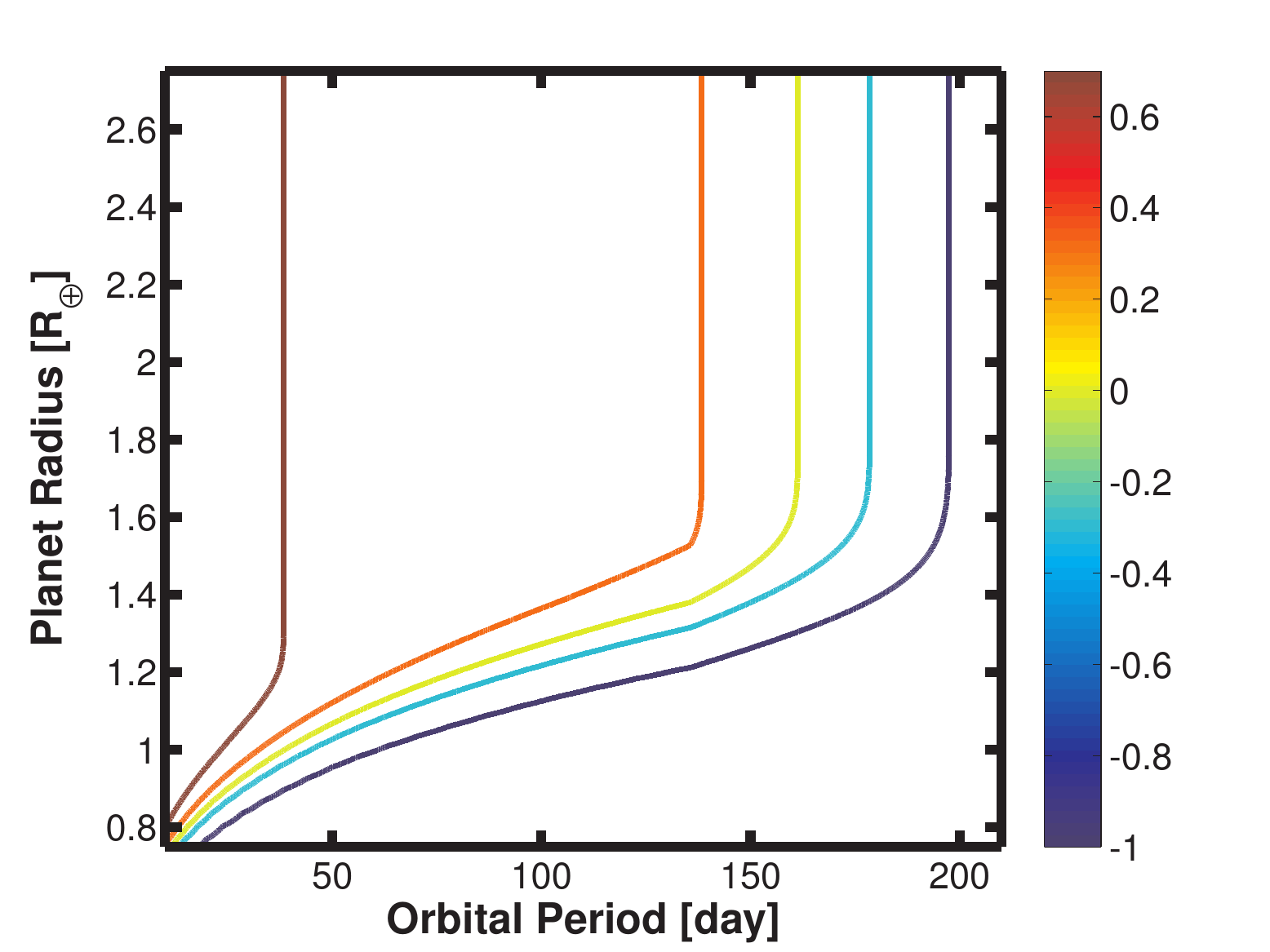}
\end{center}
\caption{The probability contours displayed in Figure~\ref{fig:completeness1} are corrected for the geometric alignment probability (normalized to the geometric alignment probability of an Earth/Sun analog: 0.00465).  These alignment-corrected probability countours are displayed here.  The colors of the contours are mapped to the logarithm of the probability.  These values, from left to right are $0.7$, $0.3$, $0.0$, $-0.3$, and $-1$.}
\label{fig:completeness2}
\end{figure}

The product of the detection probability and the probability of geometric alignment (proportional to the ratio of the stellar radius and planet-star separation normalized to 0.46\%\ for an Earth-Sun analog) yields the total detection probability.  Figure~\ref{fig:completeness2} displays (logarithmic) completeness contours for the same Kp=12 star after including the geometric alignment probability.  From left to right, the contours are $5.0$, $2.0$, $1.0$, $0.5$, and $0.1$ percent, or $0.7$, $0.3$, $0.0$, $-0.3$, and $-1$, respectively, on the logarithmic scale. If nature produces planets that uniformly populate the radius/period plane with circular orbits, then the planets detected by \ek\ will be distributed similarly to the contours shown.

Instead of computing a grid of total detection probabilities for each of the $>$150,000 stars in the parent sample, we devise a representative sample.  Stars observed by \ek, in the range $4.0<\logg<4.9$ and \rstar $< 1.4$, are sorted into Kp$=0.25$ magnitude bins, excluding those that do not have stellar classifications in the \ek\ Input Catalog.  This results in 145,728 targets with $10.75<{\rm Kp}<17.75$.  Table~\ref{tab:completeness1} lists the number of stars in each magnitude bin together with the median stellar radius and the associated \logg\ and \teff\ from interpolation of Tables $15.7$ and $15.8$ in \cite{aq}, and the photometric noise taken as the 30$^{\rm th}$ percentile of the 6-hour CDPP values of the targets in the magnitude bin.  The stellar properties associated with the Kp=12 magnitude bin are those used for the calculations that produced the results displayed in Figures~\ref{fig:completeness1} and~\ref{fig:completeness2}. Though the majority of \ek\ targets are G-type stars on or near the Main Sequence \citep{tm}, it is evident from Table~\ref{tab:completeness1} that the median star type depends on magnitude.  We assume that all stars in the magnitude bin can be represented by the properties tabulated. Though insufficient for computing {\it Kepler's} detection efficiency in an absolute sense, this assumption is a useful simplification for exploring the expected planet yield in a relative sense.

The total planet yield for a given magnitude bin is computed by taking the product of the alignment-corrected detection probability and the number of targets in that magnitude bin.  Summing these results for all magnitudes yields the total expected planet yield for a given period/radius combination assuming all stars have such a planet.  We can then integrate over period and radius to obtain the planet yields in specific bins.  These sums are weighted by the power-law distributions defined in \cite{howard} to account for the fact that certain areas of the radius/period domain are more heavily populated with planets than others.  Table~\ref{tab:completeness2} contains the expected versus observed gains in the numbers of candidates (expressed as a ratio, $N_{\rm Q6}/N_{\rm Q5}$, where N$_{\rm Q5}$ is the number of candidates in the B11 catalog and N$_{\rm Q6}$ is the total number of candidates, old plus new) for various radius/period bins.  In every case, the observed gains are significantly larger than the predicted gains.  Even in areas of the parameter space where we predict no gains (e.g. Neptune-size planets with periods shorter than 125 days where $N_{\rm Q6}/N_{\rm Q5}=1$), we observe appreciably more candidates.

This simplified exercise is meant to serve as a cautionary example for those performing statistical studies of planet occurrence rates as it demonstrates the incompleteness of the B11 catalog.  The gains we observe cannot be explained by the longer observation window.  More likely, the current batch of new planet candidates has benefitted from the growing sophistication of pipeline software, the most important of which is the multi-quarter functionality afforded by TPS for the first time in SOC 7.0.  With multi-quarter functionality, data from different quarters are stitched together within pipeline modules before whitening and combing for transits.  The analysis leading to the B11catalog employed non-pipeline tools to do similar tasks via a modified box least-squares detection method.  As such, the analysis did not make full use of the pipeline modules  that carefully treat inter-quarter boundaries and perform the pre-search whitening of the light curves as described in Section~\ref{sec:tce}.  The absence of these modules has a significant impact on the detectability of shallow transits.

\begin{figure}
\begin{center}
\includegraphics[width=175mm]{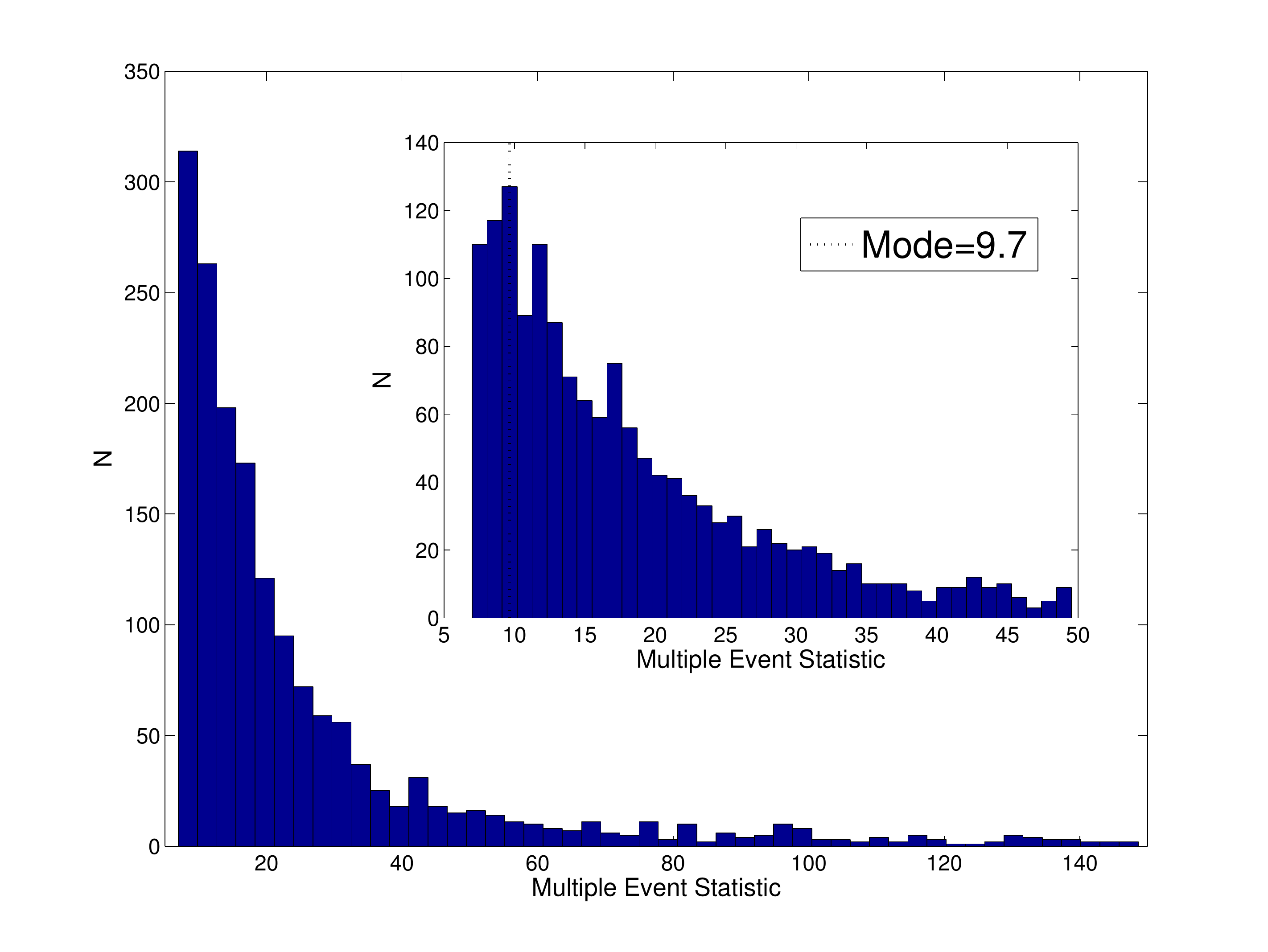}
\end{center}
\caption{Distribution of the multiple event statistic (MES) for all planet candidates.  A power law increase in the planet radius distribution toward smaller radii should translate to a power law increase in the MES distribution.  The turnover at MES=9.7 suggests that the sample is incomplete if, indeed, the inverse power law distribution of \cite{howard} holds for the smallest planets.}
\label{fig:mes}
\end{figure}

An important question is the degree to which the current catalog also suffers from incompleteness.  Figure~\ref{fig:mes} shows the distribution of Q1--Q6 detection statistics (Multiple Event Statistic) for the cumulative list of planet candidates (new plus old).  A power law increase in the planet radius distribution toward smaller radii translates to a power law increase the MES distribution (towards smaller MES values).  The turnover at a MES of 9.7 suggests that incompleteness might still be an issue.  Planned pipeline upgrades already underway are expected to yield another significant improvement.  A more detailed analysis of pipeline completeness is in progress. 

\subsection{The New Multiples}
\label{sec:multis}

The first multiple transiting planet systems were identified in \ek\ data within the first year of science operations  \citep{juneCatalog,steffen,kepler9}.  One third of the 1,235 candidates in the B11 catalog, or 17\%\ of the 997 unique stars are host to multiple planet candidates.  Evidence of dynamical interactions in the form of transit timing variations (TTV) has been identified for approximately 100 candidates \citep{fordTTV}, and dynamical modeling of those variations has successfully produced mass determinations for five of the planets associated with Kepler-11 \citep{kepler11}.   TTVs have proven useful for confirming and characterizing planets in multiple systems.  Moreover, statistical analysis of multiples indicates that nearly all ($\sim$98\%) are bona fide planets  \citep{validMulti}.  That is, the false positive rate for candidates transiting stars with multiple transiting planet candidates is significantly smaller than for single planet candidates .  As such, the ``multis'' are a particularly reliable sample for future study.    
                                                     
As discussed in Section~\ref{sec:tce}, all of the \ek\ Objects of Interest (KOIs) were inspected for evidence of additional transiting planet candidates. The search yielded 302 new candidates in multiple systems.  Their properties are included in Tables~\ref{tab:planetProps1} and~\ref{tab:planetProps2}.   Combining the new candidates with the B11 catalog, we have \nCandidatesTotal\ candidates associated with \nStarsTotal\ unique stars.  \nStarsWithMulti\ of the \nStarsTotal\ stars host multiple candidates. \nMultiCandidates\ of the \nCandidatesTotal\ candidates are part of multiple systems.  The fraction of KOI host stars with multiple candidates has risen from 17\%\ to 20\%.

\begin{figure}
\begin{center}
\includegraphics[width=175mm]{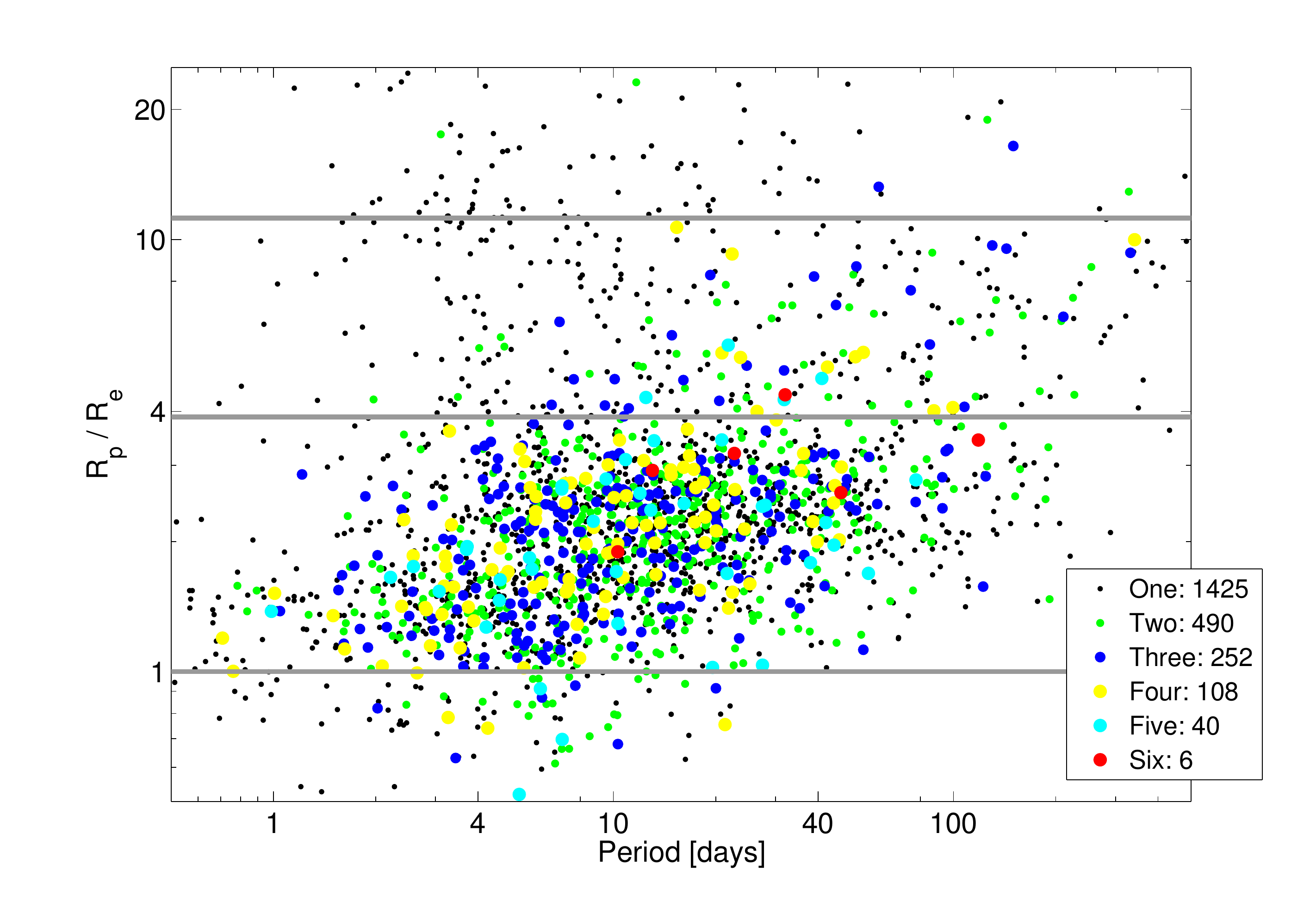}
\end{center}
\caption{Planet radius versus orbital period of the cumulative set of planet candidates as displayed in Figure~\ref{fig:radiusPeriod}.  The points are colored to display members of one, two, three, four, five, and six-planet candidate systems.  We note the continued paucity of giant planets at short orbital periods in multiple planet systems.  20\%\ of the stars catalogued have multiple planet candidates.}
\label{fig:multis}
\end{figure}

In Figure~\ref{fig:multis}, we return to the cumulative list of candidates in the radius versus period plane, this time distinguishing candidates belonging to one (black), two (green), three (blue), four (yellow), five (cyan), and six (red) planet systems.  The plot corroborates the observation made by \cite{latham} in studying the multis in the B11 catalog: that there is a paucity of short-period giant planets in multiple systems.  There is only one apparent exception to this -- the green point near $\rpl=17$\rearth\ , $P=3$~days (KOI-$338.02$).  This is a new candidate associated with KOI-$338$, the first having been reported in the B11 catalog.  Since that publication, spectroscopic observations have been acquired and subjected to spectral synthesis.  The revised stellar parameters are $\teff=4104$~K and $\logg=1.87$ compared to $\teff=4910$~K and $\logg=4.18$ in the \ek\ Input Catalog.  This results in a significant change in the estimated planet radius for KOI-338.01 reported in B11.  Re-evaluation of KOI-338.01 with the new stellar radius consistent with the lower surface gravity yields $\rpl=40 \rearth$. This transiting (eclipsing) object is too large to be planetary and is off the scale in Figure~\ref{fig:multis}.  The new candidate, KOI-$338.02$, is approximately Jupiter-size.  A second spectroscopic observation from the same instrument (the TRES spectrograph on the Tillinghast Reflector at Whipple Observatory) was acquired, and yields a surface gravity that is more consistent with the KIC value.  Additional observations are required to resolve the stellar properties.  This target is discussed in more depth by \cite{fabrycky}.

We note that Figure 1 of \cite{latham} contains other short-period giant planet candidates:   the candidates associated with KOI-$961$.  This high proper M dwarf from the proper motion catalog of \cite{lepine} (LSPM J$1928+4437$) is unclassified in the \ek\ Input Catalog.  In B11, the stellar radius was estimated by inferring an effective temperature from the J-K color and assuming a Main Sequence luminosity class. Improved stellar properties and light curve modeling of \cite{koi961} result in a considerable decrease in the star and planet radii thereby removing these candidates from the upper left corner of Figure~\ref{fig:multis} and strengthening the case for an observed paucity of short-period giant planets in multiple systems.

Application of the statistical arguments presented by \cite{validMulti} to the current population of multis implies that there are over 880 (98\%\ of 898) bona fide planets in this sample alone.  A comprehensive study of the architecture of these multis is presented by \cite{fabrycky}.

\subsection{Candidates in the Habitable Zone}
\label{sec:hz}

\cite{febCatalog} identified 54 transiting planet candidates in the Habitable Zone (HZ),  defined by an equilibrium temperature, $T_{\rm eq}$ (see Section~\ref{sec:derivedParams}), between the freezing and boiling point of water ($273$ to $373$~K).  As pointed out by \cite{kasting11}, this definition fails to take into consideration the warming effect of an atmosphere, thereby rendering many of the candidates too hot for habitability even under the most liberal assumptions about their climatic conditions.   As the roster of small planets at long orbital periods grows, so has the attention paid to the issue of habitability.   Recent modeling efforts suggest that the equilibrium temperature at the inner edge of the HZ might be closer to $270$~K \citep{selsis}, corresponding to rapid water loss via H escape or a runaway greenhouse effect, depending on the surface water content.  The outer edge depends on the fractional cloud coverage and ranges from $175$~K to $200$~K \citep{hz}.  Others have proposed defining the HZ in terms of insolation (the amount of stellar flux incident on the planet surface) in order to remove the built-in assumptions (e.g. Bond albedo) required to compute the equilibrium temperature \citep{hz2}.  

\begin{figure}
\begin{center}
\includegraphics[width=175mm]{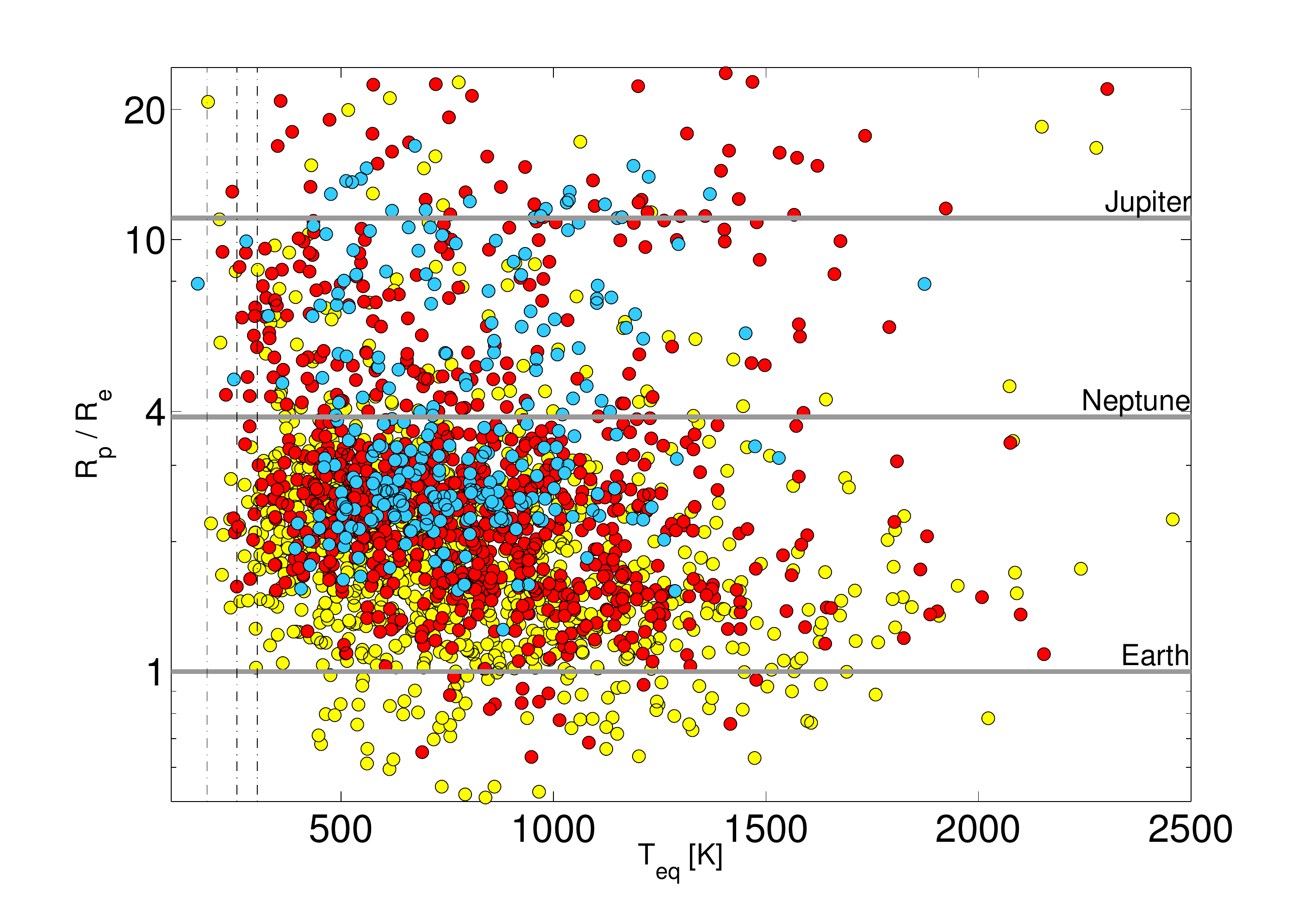}
\end{center}
\caption{Radius versus equilibrium temperature for each of the planet candidates in the B10 catalog (blue points), the B11 catalog (red points), and this contribution (yellow points).  Horizontal lines marking the radius of Jupiter, Neptune, and Earth are included for reference.  Also included for reference are vertical lines marking the inner and outer edges of the Habitable Zone as defined by \cite{hz} as well as the equilibrium temperature for an Earth-Sun analog (middle line) under the same assumptions as those described in Section~\ref{sec:hz}.}
\label{fig:radiusTeq}
\end{figure}

The star and planet properties provided in Tables~\ref{tab:starProps},~\ref{tab:planetProps1}, and~\ref{tab:planetProps2} can be used to compute the insolation and/or equilibrium temperature under different assumptions for the purpose of assessing questions of habitability on a case-by-case basis.  Here, we use equilibrium temperature as defined in Section~\ref{sec:derivedParams} to examine the population of planet candidates likely to be in or near the Habitable Zone.  Figure~\ref{fig:radiusTeq} shows planet radius versus equilibrium temperature for the entire sample of planet candidates.  For reference, we include a vertical dashed line (middle) to mark the equilibrium temperature of the Earth computed under the same set of assumptions.  With each new catalog (blue to red to yellow points), we see a clear trend toward Earth-size planets at Earth's equilibrium temperature (i.e. toward the bottom left hand corner of the diagram). 

\begin{figure}
\begin{center}
\includegraphics[width=175mm]{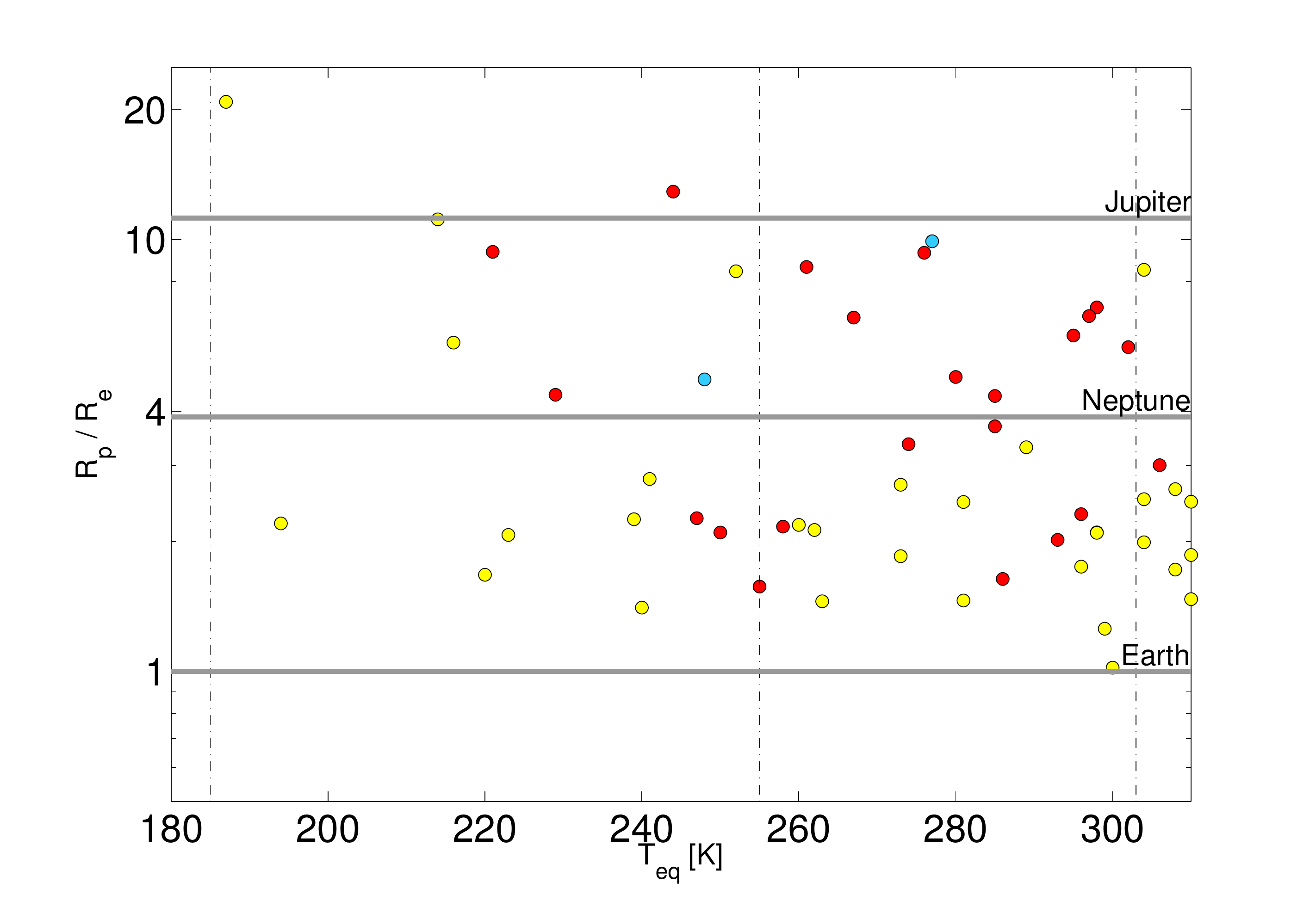}
\end{center}
\caption{Same as Figure~\ref{fig:radiusTeq} for planet candidates in and near the Habitable Zone.}
\label{fig:hzZoom}
\end{figure}

Figure~\ref{fig:hzZoom} displays the same for the range $180~{\rm K} < T_{\rm eq} < 310~{\rm K}$.  The dotted vertical lines (far left and far right) mark the (generous) HZ boundaries ($185$ K to $303$ K) proposed by  \cite{kastingExoPag}.  The intermediate dashed vertical line marks the equilibrium temperature of the Earth under the same set of assumptions.  There are \hzTotal\ candidates in this temperature range (compared to \hzTotalFeb\ in the B11 catalog). \hzSuperEarthSize\ are super-Earth-size ($1.25 \rearth \le \rpl < 2 \rearth$) and \hzEarthSize\ is Earth-size ($\rpl < 1.25  \rearth$).  Table~\ref{tab:hzTable} lists the properties for the 25 new candidates in this temperature range that are plotted in Figure~\ref{fig:hzZoom}.  Note that candidates with only one transit event in the Q1--Q6 period are excluded from this list and Figures~\ref{fig:radiusTeq} and~\ref{fig:hzZoom}.  

We have paid special attention to candidates that are in this temperature range and are also near Earth-size (see, for example, the discussion of KOI-326, KOI-364, and KOI-1026 in Appendix~\ref{sec:bigCatalog}).  Figure~\ref{fig:koi2124} shows the relative flux time-series of KOI-2124.01, the smallest viable candidate in this temperature range.

Table~\ref{tab:hzTable} contains two sets of stellar parameters.  They are identical when spectroscopic values are available (as indicated by the flag, $f_{\teff}$, in Table~\ref{tab:starProps}).  They differ where \ek\  Input Catalog values were updated using a parameter search in the Yonsei-Yale stellar evolution models as described in Section~\ref{sec:starProps}.  For the sample of stars listed in Table~\ref{tab:hzTable}, the updates almost always lead to smaller stellar radii (and, hence, smaller and cooler planet candidates). Improved stellar characterization is required for a more reliable determination of the candidate location relative to the Habitable Zone.

\subsection{Citizen Science:  Planet Hunters Discoveries}
\label{sec:planethunters}

PlanetHunters.org is a citizen science tool \citep{ph1}, based on the Zooniverse platform \citep{zooniverse}, that enables the search for transit events in the public Kepler data.  The site serves up plots of Kepler light curves broken into 30-day segments, and, through a sequence of queries, leads the user through a high level classification that sorts light curves by their qualitative properties, or appearance.  Discerning eyes flag events that resemble transits, and the goal is to have every light curve examined by at least 5 independent users.  Since its launch in December 2010, over 10 million classifications have been made by over 100,000 users, underscoring the remarkable enthusiasm of the general public.  The site affords one not only the opportunity to experience the scientific method but also the possibility of experiencing the gratification of discovery.  With its power in numbers, the citizen science project is a welcome complement to the automated detection algorithms in Kepler's software pipeline.

The Planet Hunters science team combines the results from the multiple classifications for each 30-day light curve segment to identify potential planet candidates within the publicly released Kepler data. Transit-like events flagged by the public are assessed by the Planet Hunters science team.  The Kepler project office then assists in vetting further for false prositives. This process resulted in the identification of four potential planet candidates in the Q1 public data \citep{ph1}, associated with stars KIC 10905746, KIC 8242434, KIC 6185331, and KIC 11820830, two of which (KIC 10905746 and KIC 6185331) were deemed viable candidates after inspection of vetting metrics.  These four candidates were assigned KOI numbers ($1725.01$, $1726.01$, $1727.01$, and $1728.01$, respectively), and blindly re-assessed together with the candidates identified here using Q1--Q6 data and the associated vetting metrics.  All but $1728.01$ survived as viable planet candidates.  Their properties are listed in Tables~\ref{tab:planetProps1} and~\ref{tab:planetProps2}.  KOI-1728.01 was rejected due to the large radius of the companion ($> 5$ R$_{\rm J}$) and hints of ellipsoidal variations in the light curve indicative of a higher mass (i.e. stellar) companion.  This target has been included in the Eclipsing Binary catalog \citep{eb1,eb2}.  

More recently, inspection of the Q1-Q2 pubic data by Planet Hunters led to another set of viable planet candidates associated with the stars KIC 4552729 and KIC 10005758 \citep{ph2}.  KIC 10005758 had already been assigned a KOI number due to the identification of a different transit event at a shorter period (KOI-$1783.01$).   The new, longer period candidate identified by the Planet Hunters team has been assigned the KOI number $1783.02$ while the candidate associated with KIC 4552729 has been assigned KOI number $2691.01$.  Detection statistics on KOI-$1783.02$ and KOI-$2691.01$ are not identified by the Q1-Q6 pipeline run due to a) the long orbital period (KOI-$1783.02$) and b) systematic noise sources that precluded identification of the correct orbital period ($2691.01$).  The events are readily identified in a Q1--Q8 pipeline run that more recently became available, and all vetting statistics indicate that both are strong candidates.  To maintain sample uniformity, we do not include these candidates in the tables presented here.  They will, however, be included in future catalogs.

We note that nine of the candidates presented in Table~\ref{tab:planetProps1} ($1787.01$, $1828.01$, $1858.01$, $1790.01$, $1808.01$, $1830.01$, $1613.01$, $1557.02$, $1930.04$) were independently identified by Planet Hunters (KIC 5864975, KIC 11875734, KIC 8160953, KIC 6504954, KIC 7761918, KIC 3326377, KIC 6268648, KIC 5371776, and KIC 5511081, respectively) as described by \cite{ph2}.  This further illustrates the potential for contributions from the citizen science community.

The vetting statistics employed here and applied to the Planet Hunters candidates are quickly becoming an integrated part of a mature software pipeline.  The Kepler team has worked to fine-tune these vetting metrics so that they are reliable and easily interpreted.  The objective is to eventually make them publicly available so that users interested in identifying transits can also perform the vetting that is such an integral part of identifying viable planet candidates.

\section{Summary}
\label{sec:summary}

We have analyzed pipeline results using Q1 through Q6 data.  Nearly 5,000 TCEs were evaluated.  Approximately 1,500 were identified as objects of interest.  Light curve modeling, Data Validation pipeline results, and photo-center analysis yield metrics for vetting astrophysical false positives.  The vetting process yielded nearly 700 new planet candidates.  These as well as previous candidates were subjected to a modified BLS transit detection analysis after filtering the primary transit events.  More than $300$ additional candidates associated with multiple systems were identified in this way and subjected to the same vetting metrics.  

We present \nCandidatesNewTotal\ new planet candidates and their properties (period, epoch, $\rpl/\rstar$, $d/\rstar$, and impact parameter) gleaned from light curve modeling.  Planet radius and equilibrium temperature require knowledge of the stellar properties.  Effective temperature and surface gravity from the \ek\ Input Catalog are used as input values to a Markov Chain Monte Carlo search for the best match Yonsei-Yale evolutionary model.  The result is an estimate of the stellar radius.  Forty-nine stars have a high-resolution spectrum obtained as part of the \ek\ follow-up program.  Analysis yields improved \teff, \logg, and \rstar, the latter of which comes from the Yonsei-Yale models.   Calculation of the planet radii yields \nearth, \nsuperearth, \nneptune, and \njupiter\ candidates that are Earth-size, super-Earth-size, Neptune-size, and Jupiter-size, respectively.   

The distribution in both size and orbital period of the new candidates is qualitatively similar to that of the previously published candidates, although smaller planets are more prevalent.  More than \newSmallerThanNeptune\ of the new planet candidates are smaller than Neptune (compared to \febSmallerThanNeptune\ for the B11 catalog).  The largest relative gains are seen not only for the smaller planets but also for those at longer orbital periods.  We report a growth of \gainSmall\ for candidates smaller than $2\rearth$, compared to \gainLarge\ for candidates larger than $2\rearth$, and a growth of \gainLong\ for orbital period longer than 50 days compared to an \gainShort\ increase for periods shorter than 50~days.  The gains for the smaller planets cannot be explained by the modest increase in data collection.  The observed gains exceed the computed gains even in regimes where one might expect previous catalogs to be complete (Table~\ref{tab:completeness2}).   This can be explained by improvements to the analysis pipeline, the most significant of which is the multi-quarter capability in the Transiting Planet Search and Data Validation modules.

The fraction of stars with multiple transiting planet candidates has risen from 17\%\ to 20\%.  The cumulative list of \nCandidatesTotal\ viable planet candidates contains \nStarsTotal\ unique stars, \nTwo\ of which are two-planet systems, \nThree\ of which are three-planet systems, \nFour\ of which are four-planet systems, \nFive\ of which are 5-planet systems, and one of which is a six-planet system (Kepler-11, \citealt{kepler11}).  A comparison of the single-planet systems with the multiple-planet systems shows a paucity of short-period (P$<10$~days) giant planets in multiple systems as reported by \cite{latham} and \cite{steffen12a}.

With each successive catalog, we see clear progress toward the Earth-size planets in the Habitable Zone (Figures~\ref{fig:radiusPeriod} and~\ref{fig:radiusTeq}).  Twenty-five of the new candidates are located in the range $185$ K $<$ T$_{\rm eq} < 303$ K, and one, KOI-2124.01, is near Earth-size and at the hot end of this temperature range.   The gains in the number of detections of planet candidates smaller than 2\rearth\ will be a boon for studying occurrence rates.   We proceed cautiously, however.  Assuming that planet sizes are distributed according to an inverse power law \citep{howard}, we would expect to see a similar distribution in the detection statistics of the candidates discovered to date.  The turnover in the distribution just short of the detection threshold suggests that we will see further improvements in completeness as the pipeline continues to improve.

\acknowledgements

Funding for this Discovery mission is provided by NASA's Science Mission Directorate.  This material is based on work supported by the
National Aeronautics and Space Administration under grant NNX08AR04G issued through the Kepler Participating Scientist Program.

\appendix

\section{Cumulative Catalog of Planet Candidates and Their Properties}
\label{sec:bigCatalog}

We present a cumulative catalog of planet candidates and their properties.  Candidates from the B11 catalog have been subjected to the same uniform modeling (as described in Section~\ref{sec:modeling}) using the same data set (Quarter 1 through 8) as the new candidates presented here.  Moreover, the stellar parameters of their host stars have been updated in the same manner as described in Section~\ref{sec:starProps}.  Over 300 of the host stars from previous catalogs have been observed spectroscopically as part of Kepler's ground-based follow-up program and subjected to the analyses that yield updated stellar properties used to derive the planet candidate properties.  Updated ephemerides, light curve properties, modeled light curve parameters, derived planet characteristics, and stellar properties of all planet candidates are presented in Table~\ref{tab:bigCatalog}.  Updates to the catalog include: a) cases where period aliasing has been resolved, b) cases where corrections for transit timing variations lead to improved light curve modeling, and c) new ephemerides for some candidates which were listed in B11 as having only one observed transit but which have since presented additional transits.

We note that no comprehensive effort has yet been made to remove false positives from the B11 catalog based on follow-up observations and/or new data and vetting metrics.  This will be done in a future contribution.  There are, however, a small number of exceptions.  For example, KOI-589.01 was reported in B11 as a 1.2\rearth\ candidate with a 17.5 day orbital period and a total SNR of 8.6$\sigma$.  After adding new observations, the total SNR of the signal has fallen below our detection threshold, to 3.7$\sigma$, suggesting that the event is a false alarm.  Light curve modeling did not converge.  The candidate has been omitted from Table~\ref{tab:bigCatalog}.   KOI-111.04 is a similar case.  Reported in B11 as a 2.5\rearth\ candidate in a 103.5 day orbit, this candidate now presents a total SNR of just 4.2$\sigma$.  It too, has been omitted.

Other exceptions are three candidates from B11 that are presumably small and in/near the Habitable Zone: KOI-326.01, KOI-364.01, KOI-1026.01.    KOI-326 (KIC 9880467) is unclassified in the \ek\ Input Catalog.  The radius of the planet candidate reported in the B11 catalog assumed the host star is a Main Sequence dwarf with an effective temperature defined by its $J-K$ color.  However, subsequent spectroscopic follow-up observations indicate that the host star is more likely to be a giant.  Preliminary estimates of the planet radius are larger than 3$R_{\rm J}$.  The KIC also lists an erroneous apparent magnitude in the \ek\ bandpass for this star.  The erroneous magnitude leads to a non-optimal photometric aperture precluding reliable photo-center analysis.  This will be important since there is a brighter star (KIC 9880470) less than 5\arcsec\ away.  More detailed analysis of KOI-364.01 and KOI-1026.01 suggests that the transit detection statistics  are driven by systematics in the data.  Pipeline improvements have led to a lower confidence in the planet interpretation.  Consequently, these two candidates have been removed from the cumulative catalog. All three of these candidates will be monitored closely as more data become available.

The properties listed in Table~\ref{tab:bigCatalog} are derived from the (automated, bulk) light curve modeling described in Section~\ref{sec:modeling}.  As of this writing, there are over 60 planet confirmations and characterizations in the literature based on Kepler transit detections.  Such studies involve more data products and/or specialized analysis techniques that lead to improved planet properties that we do not attempt to catalog here.  For example, KOI-1611.02 is the circumbinary planet published as Kepler-16ABb \citep{kepler16}.  Since the uniform light curve modeling does not handle the case of circumbinary systems, most of the entries in Tables~\ref{tab:planetProps1},~\ref{tab:planetProps2}, and~\ref{tab:bigCatalog} are assigned values of $-99$ indicating invalid parameters.  This is an extreme example.  However, all confirmed planets will have improved ephemerides and/or physical properties in the published literature.  The properties of confirmed planets as well as the mapping between KIC, KOI, and Kepler identification numbers can be found at the {\it NASA Exoplanet Archive}\footnote{\url{http://exoplanetarchive.ipac.caltech.edu}}.

\clearpage


\clearpage

\clearpage

\clearpage

\clearpage

\clearpage

\begin{figure}
\begin{center}
\includegraphics[width=175mm]{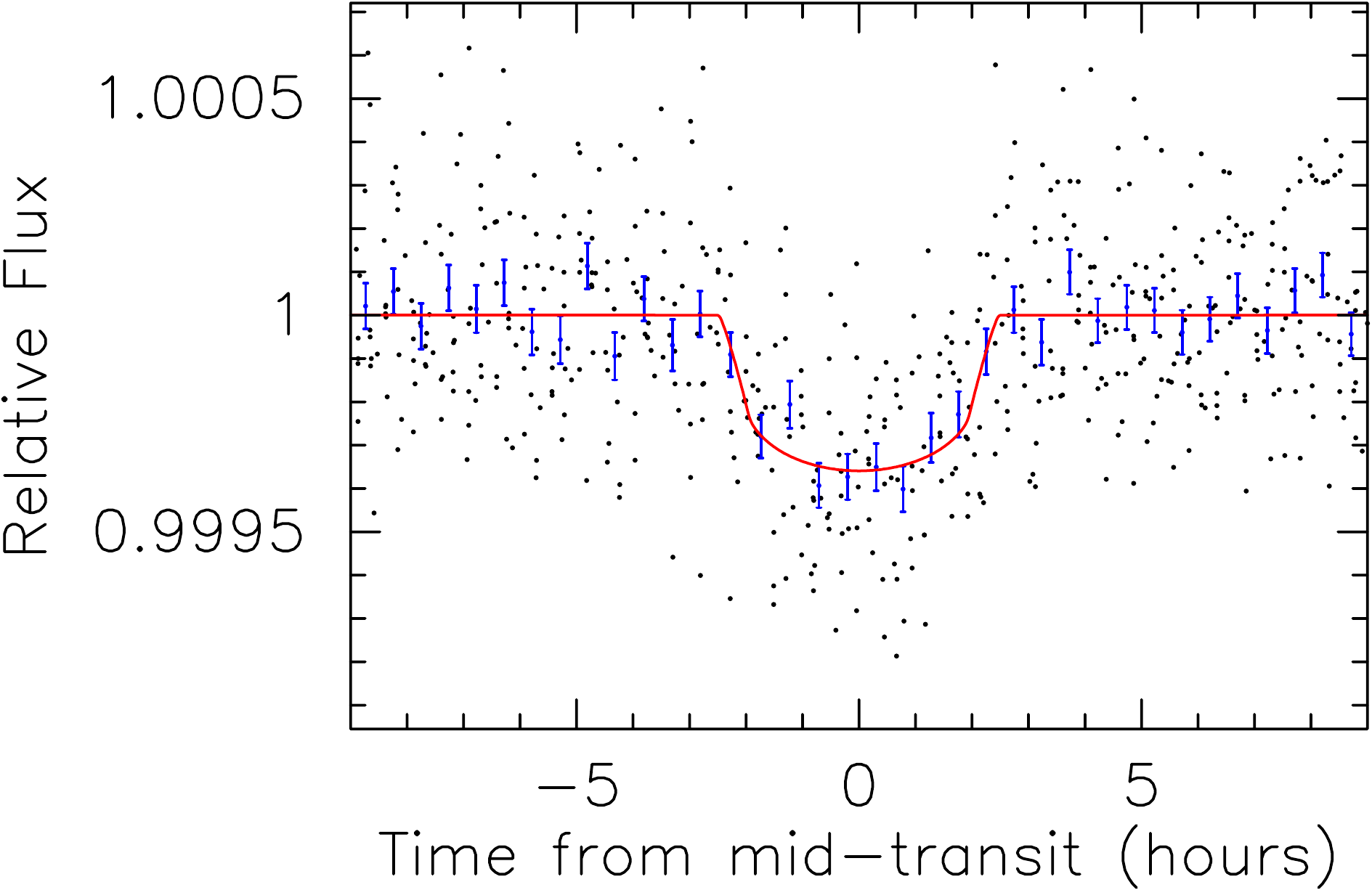}
\end{center}
\caption{Phase-folded, relative flux time-series of KOI-$2124.01$ light curve.  KOI-$2124.01$ is the smallest viable planet candidate near the HZ.}
\label{fig:koi2124}
\end{figure}
\clearpage


\begin{deluxetable}{cccccc}
\tabletypesize{\scriptsize}
\tablewidth{0pc}
\tablecaption{Data Collection Times.\label{tab:observations}}
\tablehead{
\colhead{Q} &
\multicolumn{2}{c}{First Cadence Mid-time} &
\multicolumn{2}{c}{Last Cadence Mid-time} &
\colhead{N$_{cadence}$} \\
\colhead{} &
\colhead{MJD} &
\colhead{UTC} &
\colhead{MJD} &
\colhead{UTC} & 
\colhead{}
}
\startdata
1 & 54964.011 & 2009 May 13 00:15 & 54997.481 & 2009 Jun 15 11:32 & 1639 \\
2 & 55002.0175 & 2009 Jun 20 00:25 & 55090.9649 & 2009 Sep 16 23:09 & 4354 \\
3 & 55092.7222 & 2009 Sep 18 17:19 & 55181.9966 & 2009 Dec 16 23:55 & 4370 \\
4 & 55184.8778 & 2009 Dec 19 21:04 & 55274.7038 & 2010 Mar 19 16:53 & 4397 \\
5 & 55275.9912 & 2010 Mar 20 23:47 & 55370.6600 & 2010 Jun 23 15:50 & 4633 \\
6 & 55371.9473 & 2010 Jun 24 22:44 & 55461.7939 & 2010 Sep 22 19:03 & 4397 \\
7 & 55462.6725 & 2010 Sep 23 16:08 & 55552.0491 & 2010 Dec 22 01:10 & 4375 \\
8 & 55567.8647 & 2011 Jan 06 20:45 & 55634.8460 & 2011 Mar 14 20:18 & 3279 \\
\enddata
\tablecomments{Quarters 1 through 8 are only employed in the light curve modeling used to derive the planet candidate properties.  Transits are identified on a pipeline run using Quarter 1 through 6 only.}
\tablecomments{CCD Module 3 failed in Quarter 4 on 2010 Jan 9 at $17:52$ UTC.  Targets located on module 3 during this quarter were only observed for $1,022$ cadences.}
\end{deluxetable}

\begin{deluxetable}{lccc|lccc|lccc}
\tabletypesize{\scriptsize}
\tablewidth{0pc}
\tablecaption{KOIs Noted as V-Shaped\label{tab:vshaped}}
\tablehead{
\colhead{KOI} &
\colhead{P} &
\colhead{$\frac{\rpl}{\rstar}$} &
\colhead{$1-b-\frac{\rpl}{\rstar}$} &
\colhead{KOI} &
\colhead{P} &
\colhead{$\frac{\rpl}{\rstar}$} & 
\colhead{$1-b-\frac{\rpl}{\rstar}$} &
\colhead{KOI} &
\colhead{P} &
\colhead{$\frac{\rpl}{\rstar}$} &
\colhead{$1-b-\frac{\rpl}{\rstar}$} \\
\colhead{} &
\colhead{[days]} &
\colhead{} &
\colhead{} &
\colhead{} &
\colhead{[days]} &
\colhead{} & 
\colhead{} &
\colhead{} &
\colhead{[days]} &
\colhead{} & 
\colhead{} 
}
\startdata
51.01 & 10.43 & 0.43 & -0.632 & 882.01 & 1.96 & 0.20 & -0.128 & 1773.01 & 83.10 & 0.46 & -0.783 \\
113.01 & -387 & 0.57 & -0.955 & 886.03 & 21.00 & 0.03 & -0.003 & 1783.01 & 134.48 & 0.08 & -0.008 \\
138.01 & 48.94 & 0.12 & -0.108 & 976.01 & 52.57 & 0.50 & -0.791 & 1793.01 & 3.26 & 0.30 & -0.541 \\
151.01 & 13.45 & 0.05 & -0.027 & 1020.01 & 54.36 & 0.37 & -0.623 & 1798.01 & 12.96 & 0.08 & -0.050 \\
225.01 & 0.84 & 0.43 & -0.810 & 1032.01 & -650 & 0.09 & -0.060 & 1799.01 & 1.73 & 0.47 & -0.821 \\
256.01 & 1.38 & 0.44 & -0.678 & 1095.01 & 51.60 & 0.09 & -0.020 & 1829.01 & 22.84 & 0.33 & -0.601 \\
371.01 & 498.39 & 0.30 & -0.556 & 1096.01 & -414 & 0.11 & 99.886 & 1845.02 & 5.06 & 0.29 & -0.540 \\
403.01 & 21.06 & 0.40 & -0.767 & 1118.01 & 7.37 & 0.02 & -0.011 & 1872.01 & 30.52 & 0.08 & -0.074 \\
410.01 & 7.22 & 0.36 & -0.656 & 1192.01 & -201292 & 0.32 & -0.539 & 1906.01 & 8.71 & 0.18 & -0.323 \\
417.01 & 19.19 & 0.12 & -0.106 & 1193.01 & 119.06 & 0.09 & -0.037 & 1935.01 & 15.44 & 0.14 & -0.204 \\
419.01 & 20.13 & 0.33 & -0.549 & 1209.01 & 272.07 & 0.08 & -0.009 & 1944.01 & 12.18 & 0.03 & -0.005 \\
466.01 & 9.39 & 0.08 & -0.049 & 1226.01 & 137.76 & 0.40 & -0.351 & 1968.01 & 10.09 & 0.03 & -0.006 \\
473.01 & 12.71 & 0.04 & -0.001 & 1227.01 & 2.16 & 0.31 & -0.417 & 2042.01 & 63.07 & 0.03 & -0.000 \\
601.02 & 11.68 & 0.23 & -0.421 & 1242.01 & 99.64 & 0.43 & -0.797 & 2128.01 & 24.26 & 0.24 & -0.435 \\
609.01 & 4.40 & 0.12 & -0.123 & 1359.02 & 104.82 & 0.07 & -0.014 & 2156.01 & 2.85 & 0.06 & -0.034 \\
611.01 & 3.25 & 0.10 & -0.093 & 1385.01 & 18.61 & 0.60 & -0.935 & 2189.01 & 33.36 & 0.47 & -0.901 \\
614.01 & 12.87 & 0.08 & -0.039 & 1387.01 & 23.80 & 0.44 & -0.555 & 2204.01 & 10.86 & 0.02 & -0.005 \\
617.01 & 37.87 & 0.41 & -0.720 & 1409.01 & 16.56 & 0.03 & -0.003 & 2299.01 & 16.49 & 0.10 & -0.162 \\
620.03 & 85.31 & 0.06 & -0.029 & 1426.03 & 150.03 & 0.15 & -0.196 & 2363.01 & 3.14 & 0.02 & -0.003 \\
625.01 & 38.14 & 0.18 & -0.324 & 1502.01 & 1.88 & 0.03 & -0.001 & 2370.01 & 78.73 & 0.03 & -0.019 \\
684.01 & 4.03 & 0.16 & -0.280 & 1540.01 & 1.21 & 0.38 & -0.430 & 2380.01 & 6.36 & 0.04 & -0.051 \\
698.01 & 12.72 & 0.11 & -0.027 & 1549.01 & 29.48 & 0.66 & -1.208 & 2486.01 & 4.27 & 0.02 & -0.012 \\
716.01 & 26.89 & 0.06 & -0.036 & 1560.01 & 31.57 & 0.16 & -0.220 & 2512.01 & 15.92 & 0.25 & -0.465 \\
728.01 & 7.19 & 0.10 & -0.017 & 1561.01 & 9.09 & 0.24 & -0.431 & 2513.01 & 19.01 & 0.10 & -0.183 \\
772.01 & 61.26 & 0.11 & -0.107 & 1582.01 & 186.40 & 0.08 & -0.022 & 2519.01 & 4.79 & 0.08 & -0.116 \\
797.01 & 10.18 & 0.09 & -0.013 & 1587.01 & 52.97 & 0.21 & -0.343 & 2528.01 & 12.02 & 0.05 & -0.058 \\
799.01 & 1.63 & 0.06 & -0.042 & 1591.01 & 19.66 & 0.04 & -0.006 & 2538.01 & 39.83 & 0.04 & -0.042 \\
815.01 & 34.84 & 0.35 & -0.628 & 1675.01 & 14.62 & 0.10 & -0.144 & 2572.01 & 6.38 & 0.03 & -0.001 \\
833.01 & 3.95 & 0.42 & -0.791 & 1684.01 & 62.82 & 0.06 & -0.049 & 2573.01 & 1.35 & 0.06 & -0.002 \\
838.01 & 4.86 & 0.12 & -0.121 & 1754.01 & 15.14 & 0.03 & -0.008 & 2577.01 & 18.56 & 0.20 & -0.372 \\
856.01 & 39.75 & 0.14 & -0.039 & 1761.01 & 10.13 & 0.08 & -0.110 & 2578.01 & 13.33 & 0.38 & -0.745                                                                         
\enddata
\tablecomments{Negative, integer period values are intended to flag KOIs that have presented only one transit in the Q1-Q6 data.}
\tablecomments{The complement of the impact parameter, $b$, minus the reduced planet radius, $\rpl/\rstar$, is listed in columns 4, 8, and 12.  This diagnostic serves as an indication of a grazing, or V-shaped, transit.  This value is negative if the purported planet is not fully blocking the stellar disk at mid-transit.  The closer this number is to $-2\times\frac{\rpl}{\rstar}$, the more severely it is grazing.  Grazing transit are required to model V-shaped light curves.} 
\end{deluxetable}

\begin{deluxetable}{lccccccccccccccccccccccccc}
\tabletypesize{\scriptsize}
\tablewidth{0pc}
\tablecaption{Host Star Characteristics.\label{tab:starProps}}
\tablehead{
\colhead{KOI} &
\colhead{KIC\tablenotemark{a}} &
\colhead{Kp\tablenotemark{b}} &
\colhead{CDPP\tablenotemark{c}} &
\colhead{$\alpha$ (J2000)} &
\colhead{$\delta$ (J2000)} &
\colhead{\teff} &
\colhead{\logg} &
\colhead{\rstar/\rsun} &
\colhead{\mstar/\msun\tablenotemark{d}} &
\colhead{$f_{\teff}$\tablenotemark{e}} &
\colhead{$f_{\rm Obs}$\tablenotemark{f}} \\
\colhead{} &
\colhead{} &
\colhead{} &
\colhead{[ppm]} &
\colhead{[hours]} &
\colhead{[degrees]} &
\colhead{[K]} &
\colhead{} &
\colhead{} &
\colhead{} &
\colhead{}
}
\startdata
   5 &   8554498 &  11.665 &  220.7 &  19.31598 &  44.6474 &  5861 &  4.19 &   1.42 &  1.14 & 3 &   111111 \\
  41 &   6521045 &  11.000 &  105.5 &  19.42573 &  41.9903 &  5909 &  4.30 &   1.23 &  1.11 & 3 &   111111 \\
  46 &  10905239 &  13.770 &   56.2 &  18.88370 &  48.3552 &  5764 &  4.40 &   1.10 &  1.12 & 2 &   111111 \\
  70 &   6850504 &  12.498 &  198.7 &  19.17987 &  42.3387 &  5443 &  4.45 &   0.94 &  0.90 & 3 &   111111 \\
  82 &  10187017 &  11.492 &  322.8 &  18.76552 &  47.2080 &  4908 &  4.61 &   0.74 &  0.80 & 3 &   111111 \\
  94 &   6462863 &  12.205 &   98.8 &  19.82220 &  41.8911 &  6217 &  4.33 &   1.24 &  1.20 & 2 &   100110 \\
 108 &   4914423 &  12.287 &   93.9 &  19.26564 &  40.0645 &  5975 &  4.33 &   1.21 &  1.15 & 3 &   111111
\enddata
\tablenotetext{a}{Kepler Input Catalog number.}
\tablenotetext{b}{Apparent magnitude in the Kepler bandpass.}
\tablenotetext{c}{RMS of Combined Differential Photometric Precision from Quarters 1 through 6 in units of parts per million.}
\tablenotetext{d}{Stellar Mass is derived from surface gravity and stellar radius.}
\tablenotetext{e}{Flag indicates source of \teff, \logg, and \rstar\ as follows. (0) derived using KIC J-K color and linear interpolation of luminosity class V stellar properties of Schmidt-Kaler (1982); (1) KIC \teff\ and \logg\ are used as input values for a parameter search of Yonsei-Yale evolutionary models yielding updated \teff, \logg, and \rstar; (2) \teff, \logg, and \rstar\ are derived using SPC spectral synthesis and interpolation of the Yale-Yonsei evolutionary tracks; (3) \teff, \logg, and \rstar\ are derived using SME spectral synthesis and interpolation of the Yale-Yonsei evolutionary tracks.}
\tablenotetext{f}{Concatenation of six integers, one for each of the six quarters of spacecraft data; the value of each successive integer indicates whether or not the star was observed for each of the successive quarters; a zero indicates the star was not observed that quarter; one indicates the star was observed the entire quarter; two indicates the star was observed only part of the quarter (relevant for stars on CCD Module 3 in Quarter 4); For example a value of 000111 indicates the star was observed in quarters 4, 5, and 6 but not in quarters 1, 2, or 3.}
\tablecomments{Table is published in its entirety in the electronic edition of the Astrophysical Journal. A portion is shown here for guidance regarding its form and content.}
\end{deluxetable}

\begin{deluxetable}{lcccccccccccccc}
\tabletypesize{\scriptsize}
\tablewidth{0pc}
\rotate
\tablecaption{Planet Candidate Characteristics: Light Curve Modeling.\label{tab:planetProps1}}
\tablehead{
\colhead{KOI} &
\colhead{$t_{\rm dur}$} &
\colhead{Depth} &
\colhead{SNR\tablenotemark{a}} &
\colhead{$T_{\rm 0}$\tablenotemark{b}} &
\colhead{$\sigma_{\rm T0}$} &
\colhead{Period\tablenotemark{b}} &
\colhead{$\sigma_{\rm P}$} &
\colhead{$d/\rstar$\tablenotemark{c}} &
\colhead{$\sigma_{d/\rstar}$} &
\colhead{\rpl/\rstar} &
\colhead{$\sigma_{\rpl/\rstar}$} &
\colhead{$b$\tablenotemark{d}} &
\colhead{$\sigma_{\rm b}$} &
\colhead{$\chi^2$} \\
\colhead{} &
\colhead{[hours]} &
\colhead{ppm} &
\colhead{} &
\colhead{[days]} &
\colhead{[days]} &
\colhead{[days]} &
\colhead{[days]} &
\colhead{} &
\colhead{} &
\colhead{} &
\colhead{} &
\colhead{} &
\colhead{} &
\colhead{}
}
\startdata
   5.02 &   3.6882 &     20 &    8.5 &   66.36690 &  0.01456 &    7.0518564 &  0.0002848 &    9.797375 &    0.558680 &  0.00428 &  0.00038 &  0.74970 &  0.29480 &    1.4\\
  41.02 &   4.4764 &     76 &   36.8 &   66.17580 &  0.00321 &    6.8870994 &  0.0000617 &    6.177925 &    0.120880 &  0.00918 &  0.00016 &  0.86500 &  0.10100 &    1.2\\
  41.03 &   6.1426 &     92 &   23.2 &   86.98394 &  0.00667 &   35.3331429 &  0.0006257 &   18.376942 &    0.359570 &  0.01042 &  0.00030 &  0.92010 &  0.09750 &    1.2\\
  46.02 &   3.7909 &     58 &    9.7 &   65.51465 &  0.01139 &    6.0290779 &  0.0001918 &    6.892009 &    0.118930 &  0.00799 &  0.00059 &  0.83650 &  0.22630 &    1.2\\
  70.05 &   3.6029 &     99 &   18.4 &   68.20094 &  0.00566 &   19.5778928 &  0.0002980 &   28.131385 &    3.635640 &  0.00998 &  0.00039 &  0.74920 &  0.23260 &    1.1\\
\nodata & \nodata & \nodata & \nodata & \nodata & \nodata & \nodata & \nodata & \nodata & \nodata & \nodata & \nodata & \nodata & \nodata & \nodata \\
1612.01 &   1.2288 &     30 &   12.2 &   65.68197 &  0.00209 &    2.4649988 &  0.0000198 &   11.917724 &    4.152810 &  0.00531 &  0.00025 &  0.63910 &  0.50290 &    1.3\\
1613.01 &   4.2342 &     78 &   22.0 &   74.08600 &  0.00466 &   15.8662120 &  0.0001989 &   18.014821 &    4.615850 &  0.00933 &  0.00064 &  0.78970 &  0.35850 &    1.0\\
1615.01 &   1.7195 &    108 &   31.9 &   65.23335 &  0.00161 &    1.3406380 &  0.0000107 &    4.967419 &    1.056310 &  0.00961 &  0.00024 &  0.58290 &  0.28050 &    1.6\\
1616.01 &   2.4240 &    147 &   23.0 &   72.41072 &  0.00363 &   13.9328148 &  0.0001269 &   41.632554 &   20.251940 &  0.01117 &  0.00119 &  0.35180 &  0.91970 &    1.2\\
1618.01 &   3.5409 &     29 &   19.0 &   65.25658 &  0.00455 &    2.3643203 &  0.0000365 &    3.150859 &    0.318180 &  0.00547 &  0.00020 &  0.81200 &  0.17720 &    1.2
\enddata
\tablenotetext{a}{SNR of the phase folded transit signal computed from modeling of Quarter 1 through Quarter 8 data.}
\tablenotetext{b}{Based on a linear fit to all observed transits. Periods estimated from the duration of a single transit and knowledge of the stellar radius are rounded to the nearest integer and multiplied by -1.}
\tablenotetext{c}{Ratio of the planet-star separation at the time of transit to the stellar radius. In the special case of zero eccentricity, it is the ratio of the semi-major axis to the stellar radius.}
\tablenotetext{d}{Note that there is a strong co-variance between $b$ and $d/\rstar$.}
\tablecomments{Invalid and/or missing data are given values of -99. Zero denotes a value smaller than the recorded precision.}
\tablecomments{Table is published in its entirety in the electronic edition of the Astrophysical Journal. A portion is shown here for guidance regarding its form and content.}
\end{deluxetable}

\begin{deluxetable}{lccccccccccccc}
\tabletypesize{\scriptsize}
\tablewidth{0pc}
\tablecaption{Planet Candidate Characteristics and Vetting Metrics.\label{tab:planetProps2}}
\tablehead{
\colhead{KOI} &
\colhead{Period\tablenotemark{a}} &
\colhead{\rpl\tablenotemark{b}} &
\colhead{$a$\tablenotemark{c}} &
\colhead{$T_{\rm eq}$\tablenotemark{d}} &
\colhead{$O/E_{\rm 1}$\tablenotemark{e}} &
\colhead{$O/E_{\rm 2}$\tablenotemark{f}} &
\colhead{Occ} &
\colhead{$\Delta$RA\tablenotemark{g}} &
\colhead{$\sigma_{\Delta RA}$} &
\colhead{$\Delta$DEC\tablenotemark{g}} &
\colhead{$\sigma_{\Delta DEC}$} &
\colhead{Offset\tablenotemark{h}} &
\colhead{MES\tablenotemark{i}} \\
\colhead{} &
\colhead{} &
\colhead{\rearth} &
\colhead{[AU]} &
\colhead{[K]} &
\colhead{} &
\colhead{} &
\colhead{} &
\colhead{[\arcsec]} &
\colhead{[\arcsec]} &
\colhead{[\arcsec]} &
\colhead{[\arcsec]} &
\colhead{} &
\colhead{}
}
\startdata
   5.02 &    7.0518564 &   0.66 &  0.075 &  1124 &   0.89 &    -99 &    -0.88 &   -0.67 &    3.82 &   -0.89 &    2.15 &    0.4 &    -99\\
  41.02 &    6.8870994 &   1.23 &  0.073 &  1071 &   1.73 &   0.17 &     0.73 &    0.69 &    0.75 &   -0.14 &    1.65 &    0.9 &    -99\\
  41.03 &   35.3331429 &   1.40 &  0.218 &   620 &   2.41 &    -99 &    -1.00 &   -0.58 &    0.59 &   -0.72 &    1.89 &    0.6 &    -99\\
  46.02 &    6.0290779 &   0.96 &  0.067 &  1032 &   2.60 &    -99 &    -0.19 &    0.94 &    0.63 &  -14.95 &    7.92 &    1.9 &    -99\\
  70.05 &   19.5778928 &   1.02 &  0.137 &   629 &   0.00 &    -99 &     1.95 &   -0.20 &    0.68 &    0.70 &    0.28 &    2.2 &    -99\\
  \nodata & \nodata & \nodata & \nodata & \nodata & \nodata & \nodata & \nodata & \nodata & \nodata & \nodata & \nodata & \nodata \\
1612.01 &    2.4649988 &   0.76 &  0.036 &  1606 &   1.09 &    -99 &     0.30 &     -99 &     -99 &     -99 &     -99 &    -99 &    9.3\\
1613.01 &   15.8662120 &   1.08 &  0.127 &   759 &   0.62 &   0.00 &     0.68 &   -0.32 &    0.92 &   -0.33 &    0.87 &    0.5 &   13.3\\
1615.01 &    1.3406380 &   1.17 &  0.025 &  1764 &   1.21 &   0.00 &     1.89 &   -0.08 &    0.25 &    0.33 &    0.35 &    1.0 &   16.7\\
1616.01 &   13.9328148 &   1.37 &  0.120 &   817 &   0.12 &    -99 &     1.09 &     -99 &     -99 &     -99 &     -99 &    -99 &   13.1\\
1618.01 &    2.3643203 &   0.77 &  0.037 &  1597 &   3.62 &   0.35 &    -1.75 &   -0.78 &    1.39 &   -0.15 &    0.86 &    0.6 &   10.2  
\enddata
\tablenotetext{a}{Based on a linear fit to all observed transits. For candidates with only one observed transit, the period is estimated from the duration and knowledge of the stellar radius; values are then rounded to the nearest integer and multiplied by -1.}
\tablenotetext{b}{Product of r/R* and the stellar radius given in Table 1.}
\tablenotetext{c}{Based on Newton's generalization of Kepler's third law and the stellar mass in Table~\ref{tab:starProps}.}
\tablenotetext{d}{See main text for discussion.}
\tablenotetext{e}{Odd/Even statistic derived from light curve modeling.}
\tablenotetext{f}{Odd/Even statistic reported by Data Validation pipeline.}
\tablenotetext{g}{Offset is transit source position minus target star position.}
\tablenotetext{h}{Distance to source position divided by noise.}
\tablenotetext{i}{Reported by the pre-release SOC 7.0 TPS pipeline run on Q1-Q6 data; MES is the detection statistic akin to a total SNR of the phase-folded transit but constructed using the matched filter correlation statistics over phase and period.}
\tablecomments{Invalid and/or missing data are given values of -99. Zero denotes a value smaller than the recorded precision.}
\tablecomments{Table is published in its entirety in the electronic edition of the Astrophysical Journal. A portion is shown here for guidance regarding its form and content.}
\end{deluxetable}

\begin{deluxetable}{cccccc}
\tabletypesize{\scriptsize}
\tablewidth{0pt}
\tablecaption{The parent star sample:  representative properties and star counts.\label{tab:completeness1}}
\tablehead{
\colhead{Kp} &
\colhead{\rstar} &
\colhead{\logg} &
\colhead{\teff} &
\colhead{Nstars} &
\colhead{CDPP\tablenotemark{a}}
}
\startdata 
11      &  1.144  &   4.3689  &  6055.0  &     315   & 23.3 \\
11.5   &  1.394  &   4.3440  &  6968.3  &   1282   & 27.4 \\
12      &  1.32    &  4.3462    &  6737.4  &   2327   & 31.1 \\
12.5   &  1.273  &  4.3481   &  6545.2   &  4177   & 36.9 \\
13      &  1.204  &  4.3547   &  6268.2   &  7272   & 44.5 \\
13.5   &  1.139  & 4.3705    &  6039.7   & 11892   & 55.2 \\
14      &  1.028  & 4.4303    &  5833.3   &  14961   & 69.0 \\
14.5   &  0.947  &  4.4744   &  5667.0   &  19340   & 92.0 \\
15      &  0.922  &  4.4816   &  5569.2   &   29831  & 123.3 \\
15.5   &  0.867  &  4.4856   &  5257.2    &  36818  & 168.9 \\
16      &  0.781  &  4.5081   &  4741.3    &  15900  & 221.2 \\
16.5   &  0.745  & 4.5249    &  4543.1    &  873  & 378.5 \\
17      &  0.738   &  4.5283  &  4505.5    &  603  & 532.6 \\
17.5   &  0.775  & 4.5107    &  4707.6    &  137  & 876.6 
\enddata
\tablenotetext{a}{30$^{\rm th}$ percentile of the 6-hour CDPP values (in parts-per-million) of all stars in the relevant magnitude bin.}
\end{deluxetable}

\begin{deluxetable}{cccc}
\tabletypesize{\scriptsize}
\tablewidth{0pt}
\tablecaption{Observed versus computed gains in planet candidate yield.\label{tab:completeness2}}
\tablehead{
\colhead{} &
\multicolumn{2}{c}{$1.25 \rearth < \rpl < 2 \rearth$} & 
\colhead{$2.5 \rearth < \rpl < 6 \rearth$} \\
\colhead{} &
\colhead{$5 < P < 50$ days} &
\colhead{$50 < P < 150$ days} &
\colhead{$10 < P < 125$ days}
}
\startdata
Computed\tablenotemark{a} & 1.12 & 1.35 & 1.00 \\
Observed\tablenotemark{a} &  \fracSmallShort & \fracSmallLong & \fracLarge
\enddata
\tablenotetext{a}{Gains are expressed as the ratio between the current total number of candidates (i.e. B11 catalog plus the candidates presented here) and the number of candidates from the B11 catalog.  Since the former is a product of the analysis of six quarters of data while the latter is a product of the analysis if five quarters of data, this is referred to in the text as $N_{\rm Q6}/N_{\rm Q5}$. The period and size range chosen for the rightmost column is where the B11 catalog is thought to be complete from a sensitivity standpoint (as indicated by a ratio near unity).  Note that the observed ratio is significantly larger than unity even in this domain.}
\end{deluxetable}

\begin{deluxetable}{rccccccccccccc}
\tabletypesize{\scriptsize}
\tablewidth{0pc}
\tablecaption{Planet Candidates with 185 K $< T_{\rm eq} < 303$ K.\label{tab:hzTable}}
\tablehead{
\colhead{KOI} &
\colhead{Period} &
\colhead{\rpl} &
\colhead{$T_{\rm eq}$} &
\colhead{$O/E_{\rm 1}$} &
\colhead{Offset} &
\colhead{MES} &
\colhead{SNR} &
\colhead{\teff\tablenotemark{a}} &
\colhead{\logg\tablenotemark{a}} &
\colhead{\rstar\tablenotemark{a}} &
\colhead{\teff\tablenotemark{b}} &
\colhead{\logg\tablenotemark{b}} &
\colhead{\rstar\tablenotemark{b}} \\
\colhead{} &
\colhead{[days]} &
\colhead{\rearth} &
\colhead{[K]} &
\colhead{} &
\colhead{} &
\colhead{} &
\colhead{} &
\colhead{[K]} &
\colhead{} &
\colhead{\rsun} &
\colhead{[K]} &
\colhead{} &
\colhead{\rsun}
}
\startdata
119.02 & 190.313 & 3.30 & 289 & 0.5 & 2.0 &   -99 & 63.2 & 5380 & 4.44 & 1.0 & 5380 & 4.45 & 0.9 \\
438.02 & 52.662 & 2.10 & 298 & 0.6 & 1.0 &   -99 & 26.7 & 4351 & 4.59 & 0.7 & 4351 & 4.71 & 0.6 \\
986.02 & 76.050 & 8.46 & 199 & 0.0 & 1.3 &   -99 & 15.3 & 5250 & 6.85 & 4.5 & 5250 & 6.85 & 4.5 \\
1209.01 & 272.070 & 5.77 & 216 & 0.0 & 0.6 &   -99 & 45.6 & 5316 & 4.79 & 0.6 & 5316 & 4.71 & 0.6 \\
1430.03 & 77.481 & 2.47 & 281 & 1.1 & 0.7 & 11.5 & 18.8 & 4502 & 4.60 & 0.7 & 4502 & 4.67 & 0.6 \\
1431.01 & 345.161 & 8.44 & 252 & 1.5 & 3.1 &   -99 & 164.2 & 5649 & 4.46 & 1.0 & 5649 & 4.46 & 1.0 \\
1466.01 & 281.564 & 11.13 & 214 & 2.0 & 3.2 &   -99 & 209.8 & 4768 & 4.45 & 0.9 & 4768 & 4.53 & 0.8 \\
1686.01 & 56.867 & 1.41 & 240 & 2.6 & 2.5 & 7.0 & 7.6 & 3665 & 4.47 & 0.7 & 3665 & 4.74 & 0.5 \\
1739.01 & 220.657 & 1.85 & 273 & 0.4 & 1.8 & 7.2 & 7.1 & 5677 & 4.63 & 0.8 & 5677 & 4.57 & 0.8 \\
1871.01 & 92.725 & 2.18 & 260 & 0.0 &   -99 & 20.2 & 18.3 & 4449 & 4.65 & 0.7 & 4449 & 4.68 & 0.6 \\
1876.01 & 82.532 & 2.25 & 239 & 0.6 & 0.8 & 19.8 & 33.3 & 4230 & 4.39 & 0.9 & 4230 & 4.77 & 0.5 \\
1902.01 & 137.861 & 20.83 & 187 & 0.1 & 1.3 & 18.0 & 26.2 & 3818 & 4.50 & 0.7 & 3818 & 4.73 & 0.5 \\
1938.01 & 96.915 & 2.10 & 298 & 0.5 & 1.2 & 16.7 & 40.3 & 5071 & 4.66 & 0.7 & 5071 & 4.67 & 0.7 \\
2020.01 & 110.966 & 2.07 & 223 & 0.3 & 1.0 & 14.6 & 31.8 & 4350 & 4.47 & 0.8 & 4350 & 4.77 & 0.5 \\
2102.01 & 187.746 & 2.79 & 241 & 0.9 & 0.6 & 12.6 & 18.1 & 5100 & 4.50 & 0.9 & 5100 & 4.64 & 0.6 \\
2124.01 & 42.337 & 1.02 & 300 & 0.1 & 0.6 & 12.1 & 18.0 & 4103 & 4.51 & 0.7 & 4103 & 4.73 & 0.5 \\
2290.01 & 91.502 & 1.75 & 296 & 1.1 & 1.3 & 9.9 & 17.7 & 4969 & 4.89 & 0.5 & 4969 & 4.67 & 0.7 \\
2418.01 & 86.830 & 1.67 & 220 & 1.0 & 0.9 & 8.9 & 13.5 & 3863 & 4.26 & 1.0 & 3863 & 4.74 & 0.5 \\
2469.01 & 131.190 & 2.13 & 262 & 1.5 & 0.7 & 8.5 & 17.5 & 4727 & 4.42 & 1.0 & 4727 & 4.59 & 0.7 \\
2474.01 & 176.830 & 1.45 & 263 & 0.5 & 1.8 & 8.5 & 12.1 & 5284 & 4.60 & 0.8 & 5284 & 4.60 & 0.7 \\
2626.01 & 38.098 & 1.46 & 281 & 0.2 & 2.4 & 7.2 & 10.7 & 3735 & 4.51 & 0.6 & 3735 & 4.73 & 0.5 \\
2650.01 & 34.988 & 1.26 & 299 & 0.4 & 0.6 & 7.1 & 11.4 & 3900 & 4.50 & 0.7 & 3900 & 4.74 & 0.5 \\
2770.01 & 205.383 & 2.20 & 194 & 0.4 &   -99 & 10.7 & 18.9 & 4352 & 4.64 & 0.6 & 4352 & 4.69 & 0.6 \\
2841.01 & 159.391 & 2.70 & 273 & 0.4 &   -99 & 8.8 & 13.2 & 5213 & 4.64 & 0.8 & 5213 & 4.60 & 0.8 \\
\enddata
\tablenotetext{a}{Original values from the Kepler Input Catalog are displayed here unless spectroscopic stellar parameters are available (as indicated by the column labeled f$_{\teff}$ in Table~\ref{tab:starProps}).}
\tablenotetext{b}{Values reproduced from Table~\ref{tab:starProps}.  Here, Kepler Input Catalog values are updated using the Yonsei-Yale evolutionary tracks as described in Section~\ref{sec:starProps}. Planet radius and equilibrium temperature are derived using these stellar properties.}
\tablecomments{Columns 1 through 8 are defined in Tables~\ref{tab:planetProps1} and~\ref{tab:planetProps2}}
\tablecomments{Unavailable or invalid entries are assigned -99}
\end{deluxetable}

\begin{deluxetable}{llll}
\tabletypesize{\scriptsize}
\tablewidth{0pc}
\tablecaption{Description of Cumulative Planet Candidate Catalog\label{tab:bigCatalog}}
\tablehead{
\colhead{Column} &
\colhead{Format} &
\colhead{Name} &
\colhead{Description}
}
\startdata
1  &  F7.2       & KOI & Kepler Object of Interest number \\
2  &  I9            & KIC & Kepler Input Catalog Identifier \\
3  &  F7.3      & Kp & Kepler magnitude \\
4  &  F10.5   & $T_0$\tablenotemark{a} & Time of a transit center; BJD-2454900 \\
5  &  F8.5     & $\sigma_{T0}$ & Uncertainty in T$_0$ \\
6  &  F12.7   & Period\tablenotemark{a} & Average interval between transits in days \\
7  &  F10.7   & $\sigma_{P}$ & Uncertainty in period \\
8  &  F6.2     & \rpl\tablenotemark{b} & Planetary radius in Earth radii=6378 km \\
9  &  F6.3     & a\tablenotemark{c} & Semi-major axis of orbit \\
10 & I5          & $T_{eq}$\tablenotemark{d} & Equilibrium temperature of planet \\
11 & F8.4     & $t_dur$ & Transit duration, first contact to last contact, in hours \\
12 & I6          & Depth & Transit depth at center of transit in parts per million \\
13 & F11.6    & $d/\rstar$\tablenotemark{e} & Ratio of planet-star separation to stellar radius \\
14 & F11.6    & $\sigma_{d/\rstar}$ & Uncertainty in $d/R*$ \\
15 & F8.5      & \rpl/\rstar & Ratio of planet radius to stellar radius \\
16 & F8.5      & $\sigma_{\rpl/\rstar}$ & Uncertainty in $r/R*$ \\
17 & F8.5       & $b$\tablenotemark{f} & Impact parameter of transit \\
18 & F8.5       & $\sigma_b$ & Uncertainty in $b$ \\
19 & F6.1       & SNR\tablenotemark{g} & Total SNR of all transits detected \\
20 & F5.2      & $\chi^2$ & Goodness of fit metric \\
21 & I5           & \teff & Stellar effective temperature \\
22 & F5.2      & \logg & Log of stellar surface gravity \\
23 & F6.2      & \rstar/\rsun & Stellar Radius \\
24 & I1          & f$_{\teff}$\tablenotemark{h} & Flag on Teff
\enddata
\tablenotetext{a}{Based on a linear fit to all observed transits. For candidates with only one observed transit, the period is estimated from the duration and knowledge of the stellar radius; values are then rounded to the nearest integer and multiplied by -1.}
\tablenotetext{b}{Product of $r/R*$ and the stellar radius given in Table 1.}
\tablenotetext{c}{Based on Newton's generalization of Kepler's third law and the stellar mass computed from surface gravity and stellar radius and given in Table~\ref{tab:starProps}.}
\tablenotetext{d}{See main text for discussion.}
\tablenotetext{e}{Ratio of the planet-star separation at the time of transit to the stellar radius. In the special case of zero eccentricity, it is the ratio of the semi-major axis to the stellar radius.}
\tablenotetext{f}{Note that there is a strong co-variance between $b$ and $d/\rstar$.}
\tablenotetext{g}{SNR of the phase folded transit signal computed from modeling of Quarter 1 through Quarter 8 data.}
\tablenotetext{h}{Flag indicates source of \teff, \logg, and \rstar\ as follows: (0) derived using KIC J-K color and linear interpolation of luminosity class V stellar properties of Schmidt-Kaler (1982); (1) KIC \teff\ and \logg\ are used as input values for a parameter search of Yonsei-Yale evolutionary models yielding updated \teff, \logg, and \rstar; (2) \teff, \logg, and \rstar\ are derived using SPC spectral synthesis and interpolation of the Yale-Yonsei evolutionary tracks; (3) \teff, \logg, and \rstar\ are derived using SME spectral synthesis and interpolation of the Yale-Yonsei evolutionary tracks.}
\tablecomments{No attempt has been made here to remove false-positives from the table of \cite{febCatalog}.  Uniform vetting of these earlier KOIs is in progress.}
\tablecomments{The table described here is published in its entirety in the electronic edition of the Astrophysical Journal.}
\end{deluxetable}

\end{document}